\documentclass[aps,a4paper,superscriptaddress,preprintnumbers,showpacs,amsmath,amssymb]{revtex4}
\usepackage{graphicx}
\usepackage{dcolumn}
\usepackage{color}
\usepackage{latexsym,amsfonts}
\usepackage{textcomp}
\usepackage{bm}

\usepackage{ulem}

\begin{document}

 \title{Precision Theoretical Analysis of Neutron Radiative Beta Decay
 }

 \author{A. N. Ivanov}\email{ivanov@kph.tuwien.ac.at}
 \affiliation{Atominstitut, Technische Universit\"at Wien,
   Stadionallee 2, A-1020 Wien, Austria}
 \author{R.~H\"ollwieser}\email{roman.hoellwieser@gmail.com}
 \affiliation{Atominstitut, Technische Universit\"at Wien,
   Stadionallee 2, A-1020 Wien, Austria}\affiliation{Department of
   Physics, New Mexico State University, Las Cruces, New Mexico 88003,
   USA} \author{N. I. Troitskaya}\email{natroitskaya@yandex.ru}
 \affiliation{Atominstitut, Technische Universit\"at Wien,
   Stadionallee 2, A-1020 Wien, Austria}
 \author{M. Wellenzohn}\email{max.wellenzohn@gmail.com}\affiliation{Atominstitut,
   Technische Universit\"at Wien, Stadionallee 2, A-1020 Wien,
   Austria}\affiliation{FH Campus Wien, University of Applied
   Sciences, Favoritenstra\ss e 226, 1100 Wien, Austria}
 \author{Ya. A. Berdnikov}\email{berdnikov@spbstu.ru}\affiliation{
   Peter the Great St. Petersburg Polytechnic University,
   Polytechnicheskaya 29, 195251, Russian Federation}

\date{\today}

\begin{abstract}
In the Standard Model of electroweak interactions and in the
tree--approximation we calculate the rate and branching ratio of the
neutron radiative $\beta^-$--decay with one--real photon emission by
taking into account the contributions of the weak magnetism and proton
recoil to order $1/m_p$ of the large proton mass $m_p$ expansion. We
find that the obtained contributions of the weak magnetism and proton
recoil increase the rate and branching ratio of the neutron radiative
$\beta^-$--decay by about $0.70\,\%$. This is large compared with the
contribution of the weak magnetism and proton recoil of about
$0.16\,\%$ to the rate of the neutron $\beta^-$--decay, calculated in
Phys. Rev. D {\bf 88}, 073002 (2013).
\end{abstract}
\pacs{12.15.Ff,
13.15.+g, 23.40.Bw, 26.65.+t}
\maketitle

\section{Introduction}
\label{sec:introduction}

For a long time the radiative $\beta^-$--decay of a free neutron $n
\to p + e^- + \bar{\nu}_e + \gamma$ was used in the analysis of the
radiative corrections to the neutron $\beta^-$--decay for the
cancellation of the infrared divergences, coming from the one--virtual
photon exchanges \cite{Berman1958}--\cite{Shann1971}. For the first
time to treat the radiative $\beta^-$--decay of a free neutron as a
physical process was proposed by Gaponov and Khafizov
\cite{Gaponov1996}, who made the first calculation of the energy
spectrum and decay rate. Then, the radiative $\beta^-$--decay was
reinvestigated in \cite{Bernard2004} and \cite{Ivanov2013}. The
branching ratio ${\rm BR}_{\beta \gamma} = 2.87 \times 10^{-3}$,
calculated for the lifetime of the neutron $\tau_n = 879.6(1.1)\,{\rm
  s}$ in \cite{Ivanov2013,Ivanov2013a} for the photon energy region
$15\,{\rm keV} \le \omega \le 340\,{\rm keV}$, agrees well within one
standard deviation with the first experimental values ${\rm BR}_{\beta
  \gamma} = 3.13(35)\times 10^{-3}$ and ${\rm BR}_{\beta \gamma} =
3.09(32)\times 10^{-3}$, measured by Nico {\it et al.} \cite{Nico2006}
and Cooper {\it et al.}  \cite{Cooper2010}, respectively. The
experimental values agree also well with the result ${\rm BR}_{\beta
  \gamma} = 2.85\times 10^{-3}$, calculated by Gardner \cite{Nico2006}
using the theoretical decay rate, published in \cite{Bernard2004}. The
analysis of the $T$--odd momentum correlations even to order
$O(\alpha^2)$ and the CP--violation by interactions beyond the
Standard Model in the neutron radiative $\beta^-$--decay has been
performed by Gardner and He in \cite{Gardner2012} and
\cite{Gardner2013}, respectively. Recently the new precise
experimental values of the branching ratios of the radiative
$\beta^-$--decay of a free neutron have been reported by the RDK II
Collaboration Bales {\it et al.} \cite{Bales2016}: ${\rm BR}_{\beta
  \gamma} = 0.00335 \pm 0.00005[\rm stat]\pm 0.00015[\rm syst]$ and
${\rm BR}_{\beta \gamma} = 0.00582 \pm 0.00023[\rm stat]\pm
0.00062[\rm syst]$, measured for the photon energy regions $14\,{\rm
  keV} \le \omega \le 782\,{\rm keV}$ and $0.4\,{\rm keV} \le \omega
\le 14\,{\rm keV}$, respectively.

This paper is addressed to the calculation of the rate of the neutron
radiative $\beta^-$--decay $n \to p + e^- + \bar{\nu}_e + \gamma$ in
the Standard Model and in the tree--approximation by taking into
account the contributions of the weak magnetism and proton recoil to
order $1/M$, where $M = (m_n + m_p)/2$ is the averaged nucleon
mass. The latter is important for a calculation of a robust
theoretical background for the experimental analysis of interactions
beyond the Standard Model \cite{Ivanov2013,Ivanov2013a}. The paper is
organized as follows. In section \ref{sec:first} we give the
analytical expressions for the amplitude and rate of the neutron
radiative $\beta^-$-decay. In section \ref{sec:conclusion} in Table I
we adduce the results of the numerical analysis of the branching
ratios of the neutron radiative $\beta^-$--decay for the experimental
regions of photons energies. We discuss the obtained results and
perspectives of the further theoretical analysis of the neutron
radiative $\beta^-$--decay. In the Appendix we give the detailed
calculation of the amplitude and rate of the neutron radiative
$\beta^-$--decay in the tree--approximation.

\section{Radiative $\beta^-$--decay of neutron in the tree--approximation to next--to--leading
 order in the large $M$ expansion}
\label{sec:first}

In the Standard Model the neutron radiative $\beta^-$--decay is
described by the following interactions
\begin{eqnarray}\label{eq:1}
\hspace{-0.3in}{\cal L}_{\rm int}(x) = {\cal L}_{\rm W}(x) + {\cal
  L}_{\rm em}(x).
\end{eqnarray}
Here ${\cal L}_{\rm W}(x)$ is the effective Lagrangian of low--energy
$V-A$ interactions with a real axial coupling constant $\lambda =
-1.2750(9)$ \cite{Abele2008} and the contribution of the weak
magnetism \cite{Ivanov2013}
\begin{eqnarray}\label{eq:2}
\hspace{-0.3in}{\cal L}_{\rm W}(x) = -
\frac{G_F}{\sqrt{2}}\,V_{ud}\,\Big\{[\bar{\psi}_p(x)\gamma_{\mu}(1 +
  \lambda \gamma^5)\psi_n(x)] + \frac{\kappa}{2 M}
\partial^{\nu}[\bar{\psi}_p(x)\sigma_{\mu\nu}\psi_n(x)]\Big\}
        [\bar{\psi}_e(x)\gamma^{\mu}(1 - \gamma^5)\psi_{\nu}(x)],
\end{eqnarray}
invariant under time reversal, where $G_F = 1.1664 \times
10^{-11}\,{\rm MeV^{-2}}$ is the Fermi coupling constant, $|V_{ud}| =
0.97417(21)$ is the Cabibbo--Kobayashi--Maskawa matrix element and
$\kappa = \kappa_p - \kappa_n = 3.7058$ is the isovector anomalous
magnetic moment of the nucleon, defined by the anomalous magnetic
moments of the proton $\kappa_p = 1.7928$ and the neutron $\kappa_n =
- 1.9130$ and measured in nuclear magneton \cite{PDG2016}. Then,
$\psi_p(x)$, $\psi_n(x)$, $\psi_e(x)$ and $\psi_{\nu}(x)$ are the
field operators of the proton, neutron, electron and antineutrino,
respectively, $\gamma^{\mu}$, $\gamma^5$ and $\sigma^{\mu\nu} =
\frac{i}{2}(\gamma^{\mu}\gamma^{\nu} - \gamma^{\nu}\gamma^{\mu})$ are
the Dirac matrices \cite{IZ1980}. Then, ${\cal L}_{\rm em}(x)$ is the
Lagrangian of the electromagnetic interaction
\begin{eqnarray}\label{eq:3}
\hspace{-0.3in}{\cal L}_{\rm em}(x) = -
e\Big\{[\bar{\psi}_p(x)\gamma_{\mu}\psi_p(x)] -
[\bar{\psi}_e(x)\gamma^{\mu}\psi_e(x)]\Big\} A_{\mu}(x),
\end{eqnarray}
where $e$ is the proton electric charge, related to the
fine--structure constant $\alpha$ by $e^2 = 4\pi \alpha$, and
$A_{\mu}(x)$ is the electromagnetic potential \cite{IZ1980}.

\begin{figure}
\centering \includegraphics[height=0.09\textheight]{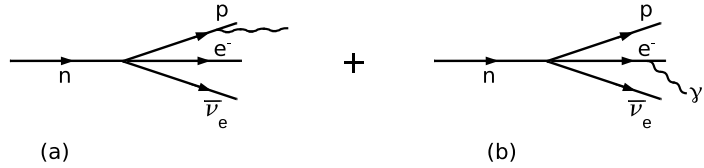}
  \caption{Feynman diagrams, defining the amplitude of the neutron
    radiative $\beta^-$--decay in the tree--approximation.}
\label{fig:fig1}
\end{figure}

In the tree--approximation the Feynman diagrams of the amplitude of
the neutron radiative $\beta^-$--decay are shown in
Fig.\,\ref{fig:fig1}.
The amplitude of the neutron radiative $\beta^-$--decay, defined by
the diagrams in Fig.\,\ref{fig:fig1}, we describe by the expression
\cite{Ivanov2013}
\begin{eqnarray}\label{eq:4}
\hspace{-0.3in}&&M(n \to p e^- \bar{\nu}_e \gamma)_{\lambda'} =
e\,\frac{G_F}{\sqrt{2}}\,V_{ud}\,{\cal M}(n \to p e^- \bar{\nu}_e
\gamma)_{\lambda'},
\end{eqnarray}
where ${\cal M}(n \to p e^- \bar{\nu}_e \gamma)_{\lambda'}$ is equal
to
\begin{eqnarray}\label{eq:5}
\hspace{-0.3in}{\cal M}(n \to p e^- \bar{\nu}_e \gamma)_{\lambda'} &=&
\Big[\bar{u}_p(\vec{k}_p, \sigma_p)
  \,\hat{\varepsilon}^*_{\lambda'}(k)\,\frac{1}{m_p - \hat{k}_p -
    \hat{k} - i0}\,O_{\mu} u_n(\vec{k}_n,
  \sigma_n)\Big]\Big[\bar{u}_e(\vec{k}_e,\sigma_e) \gamma^{\mu} (1 -
  \gamma^5) v_{\nu}(\vec{k}, + \frac{1}{2})\Big]\nonumber\\
\hspace{-0.3in}&-& \Big[\bar{u}_p(\vec{k}_p, \sigma_p) O_{\mu}
  u_n(\vec{k}_n, \sigma_n)\Big]
\Big[\bar{u}_e(\vec{k}_e,\sigma_e)\,\hat{\varepsilon}^*_{\lambda'}(k)\,
  \frac{1}{m_e - \hat{k}_e - \hat{k} - i0}\, \gamma^{\mu} (1 -
  \gamma^5) v_{\nu}(\vec{k}, + \frac{1}{2})\Big].
\end{eqnarray}
The matrix $O_{\mu}$ is given by \cite{Ivanov2013}
\begin{eqnarray}\label{eq:6} 
O_{\mu} = \gamma_{\mu}(1 + \lambda \gamma^5) + i\,\frac{\kappa}{2
  M}\,\sigma_{\mu\nu}(k_p - k_n)^{\nu}.
\end{eqnarray}
Then, $\bar{u}_p(\vec{k}_p, \sigma_p)$, $ u_n(\vec{k}_n, \sigma_n)$,
$\bar{u}_e(\vec{k}_e,\sigma_e)$ and $v_{\nu}(\vec{k}_{\nu}, +
\frac{1}{2})$ are the Dirac wave functions of the free proton,
neutron, electron and electron antineutrino with 3--momenta
$\vec{k}_p$, $\vec{k}_n = \vec{0}$, $\vec{k}_e$ and $\vec{k}_{\nu}$
and polarizations $\sigma_p = \pm 1$, $\sigma_n = \pm 1$, $\sigma_e =
\pm 1$ and $+ \frac{1}{2}$ \cite{Ivanov2013,Ivanov2014}, respectively,
$\varepsilon^{*\alpha}_{\lambda'}(k)$ is the polarization vector of
the photon in the polarization state $\lambda' = 1,2$ with a
4--momentum $k$, obeying the constraint
$\varepsilon^{*}_{\lambda'}(k)\cdot k = 0$. The amplitude
Eq.(\ref{eq:5}) is gauge invariant. Indeed, one may show that,
replacing $\varepsilon^{\alpha *}_{\lambda'}(k) \to k^{\alpha}$ and
using the Dirac equations for the free proton and electron, the
amplitude Eq.(\ref{eq:5}) vanishes.

The rate of the neutron radiative $\beta^-$--decay, described by the
Feynman diagrams in Fig.\,\ref{fig:fig1} with photon from the energy
region $\omega_{\rm min} \le \omega \le \omega_{\rm max}$, is equal to
(see \cite{Ivanov2013} and Eq.(\ref{eq:A.33}) of the Appendix)
\begin{eqnarray}\label{eq:7}
\hspace{-0.3in}&&\lambda_{\beta \gamma}(\omega_{\rm max},\omega_{\rm
  min})= (1 + 3 \lambda^2)\frac{\alpha}{\pi} \frac{G^2_F
  |V_{ud}|^2}{2\pi^3} \int^{\omega_{\rm max}}_{\omega_{\rm
    min}}\!\!\frac{d\omega}{\omega}\int^{E_0 - \omega}_{m_e}\!\!\!\!\!\!\!
dE_e\,E_e\sqrt{E^2_e - m^2_e}\, (E_0 - E_e - \omega)^2\,F(E_e, Z =
1)\,\rho_{\beta\gamma}(E_e,\omega),\nonumber\\
\hspace{-0.3in}&&
\end{eqnarray}
where $E_0 = (m^2_n - m^2_p + m^2_e)/2 m_n$ is the end--point energy
of the electron energy--spectrum of the neutron $\beta^-$--decay
\cite{Ivanov2013} and $F(E_e, Z = 1)$ is the relativistic Fermi
function, describing Coulomb proton--electron final--state
interaction. It is equal to \cite{Ivanov2013}
\begin{eqnarray}\label{eq:8}
\hspace{-0.3in}F(E_e, Z = 1 ) =  \Big(1 +
\frac{1}{2}\gamma\Big)\,\frac{4(2 r_pm_e\beta)^{2\gamma}}{\Gamma^2(3 +
  2\gamma)}\,\frac{\displaystyle e^{\,\pi
 \alpha/\beta}}{(1 - \beta^2)^{\gamma}}\,\Big|\Gamma\Big(1 + \gamma +
 i\,\frac{\alpha }{\beta}\Big)\Big|^2,
\end{eqnarray}
where $\beta = k_e/E_e = \sqrt{E^2_e - m^2_e}/E_e$ is the electron
velocity, $\gamma = \sqrt{1 - \alpha^2} - 1$, $r_p$ is the electric
radius of the proton and $\alpha = 1/137.036$ is the fine--structure
constant.  In numerical calculations we will use $r_p = 0.841\,{\rm
  fm}$ \cite{LEP}. The function $\rho_{\beta \gamma}(E_e,\omega)$ is
calculated in the Appendix. It is given by the integral
\begin{eqnarray}\label{eq:9}
\hspace{-0.3in}&&\rho_{\beta\gamma}(E_e,\omega) =
\int\frac{d\Omega_{e\gamma}}{4\pi}\,\Big[1 +
  2\,\frac{\omega}{M}\,\frac{E_e - \vec{k}_e\cdot \vec{n}}{E_0 - E_e -
    \omega} + \frac{3}{M}\,\Big(E_e + \omega - \frac{1}{3}\,E_0\Big) +
  \frac{\lambda^2 - 2(\kappa + 1)\lambda + 1}{1 +
    3\lambda^2}\,\frac{E_0 - E_e - \omega}{M}\Big]\nonumber\\
\hspace{-0.3in}&&\times\,\Big[\Big(1 +
  \frac{\omega}{E_e}\Big)\,\frac{k^2_e - (\vec{k}_e\cdot
    \vec{n}\,)^2}{(E_e - \vec{k}_e\cdot \vec{n}\,)^2} +
  \frac{\omega^2}{E_e}\,\frac{1}{E_e - \vec{k}_e\cdot \vec{n}}\Big] +
\frac{3\lambda^2 - 1}{1 + 3 \lambda^2}\,\frac{1}{M}\,\Big\{\frac{k^2_e + \omega \vec{k}_e\cdot
  \vec{n}}{E_e}\,\Big(\frac{k^2_e - (\vec{k}_e\cdot \vec{n}\,)^2}{(E_e
  - \vec{k}_e\cdot \vec{n}\,)^2} + \frac{\omega}{E_e - \vec{k}_e\cdot
  \vec{n}}\Big)\nonumber\\
\hspace{-0.3in}&& + (\omega + \vec{k}_e\cdot \vec{n}\,)\Big[\Big(1 +
  \frac{\omega}{E_e}\Big)\frac{\omega}{E_e - \vec{k}_e\cdot \vec{n}} -
  \frac{m^2_e}{E_e}\,\frac{\omega}{(E_e - \vec{k}_e\cdot
    \vec{n}\,)^2}\Big]\Big\} - \frac{\lambda^2 + 2 (\kappa + 1)\lambda
  - 1}{1 + 3\lambda^2}\,\frac{1}{M}\,\Big[\frac{k^2_e + \omega^2 +
    2\omega \vec{k}_e\cdot \vec{n}}{E_e}\nonumber\\
\hspace{-0.3in}&&\times\,\frac{k^2_e - (\vec{k}_e\cdot
  \vec{n}\,)^2}{(E_e - \vec{k}_e\cdot \vec{n}\,)^2} +
\frac{\omega}{E_e}\,\frac{k^2_e - (\vec{k}_e\cdot \vec{n}\,)^2}{E_e -
  \vec{k}_e\cdot \vec{n}} + \frac{\omega^2}{E_e}\,\frac{\omega +
  \vec{k}_e\cdot \vec{n}}{E_e - \vec{k}_e\cdot \vec{n}}\Big] -
\frac{\lambda(\lambda - 1)}{1 +
  3\lambda^2}\,\frac{1}{M}\,\Big[\frac{\omega}{E_e}\, \frac{k^2_e -
    (\vec{k}_e\cdot \vec{n}\,)^2}{E_e - \vec{k}_e\cdot \vec{n}} +
  3\,\frac{\omega^2}{E_e}\Big].
\end{eqnarray}
Here $d\Omega_{e\gamma}$ is an infinitesimal solid angle of the
electron--photon momentum correlations $\vec{k}_e\cdot \vec{n} = k_e
\cos\theta_{e\gamma}$, where $\vec{n} = \vec{k}/\omega$ is a unit
vector along the photon 3--momentum \cite{Ivanov2013,Ivanov2013a}. The
results of the numerical analysis of the rate of the neutron radiative
$\beta^-$--decay, calculated relative to the neutron lifetime $\tau_n
= 879.6(1.1)\,{\rm s}$ \cite{Ivanov2013,Ivanov2013a}, we give and
discuss in section \ref{sec:conclusion}.

\section{Discussion and Conclusion}
\label{sec:conclusion}

Recent new experimental measurements of the branching ratio of the
neutron radiative $\beta^-$--decay, reported by the RDK II
Collaboration Bales {\it et al.} \cite{Bales2016}, have been the
impetus for our theoretical analysis of the neutron radiative
$\beta^-$--decay. We have performed the calculation of the amplitude
of the neutron radiative $\beta^-$--decay in the tree--approximation
to next--to--leading order in the large proton mass expansion. We have
taken into account the complete set of contributions of the weak
magnetism and proton recoil to order $1/M$, where $M$ is the averaged
nucleon mass. The obtained results we consider as a first step towards
the precision theoretical analysis of the neutron radiative
$\beta^-$--decay in the Standard Model to a relative order $10^{-3}$
\cite{Ivanov2013,Ivanov2013a}.  The detailed calculations of the
amplitude and decay rate we give in the Appendix. The numerical values
of the branching ratios of the neutron radiative $\beta^-$--decay,
calculated relative to the neutron lifetime $\tau_n = 879.6(1.1)\,{\rm
  s}$ \cite{Ivanov2013,Ivanov2013a}, for the experimental regions of
photon energies are adduced in the Table\,I. 

\begin{table}[h]
\begin{tabular}{|c|c|c|c|c|}
\hline $\omega\, [\rm keV]$ & ${\rm BR}_{\beta\gamma} (\rm
Experiment)$ & ${\rm BR}_{\beta\gamma} (\rm Theory)$ & ${\rm
  BR}_{\beta\gamma}\, (\rm Theory)\, M \to \infty$ & $ \Delta{\rm
  BR}_{\beta\gamma}\,(\rm Theory)$\\ \hline $15 \le \omega \le 340$ &
$~~~~~~~~~~~~~~(3.09 \pm 0.32)\times 10^{-3}$
\quad~~~~~~\cite{Cooper2010} & $2.89 \times 10^{-3}$ & $2.87 \times
10^{-3}$& $0.70\,\%$\\\hline $14 \le \omega \le 782$ & $(3.35 \pm
0.05\,[\rm stat] \pm 0.15\,[syst])\times 10^{-3}$\;\cite{Bales2016} &
$ 3.04 \times 10^{-3}$ & $3.02 \times 10^{-3}$ & $0.66\,\%$ \\ \hline
$0.4 \le \omega \le 14$ & $(5.82 \pm 0.23\,[\rm stat]\pm
0.62\,[syst])\times 10^{-3}$\;\cite{Bales2016} & $5.08 \times 10^{-3}$
& $5.05 \times 10^{-3}$ & $0.60\,\%$\\\hline
\end{tabular} 
\caption{Branching ratios of the radiative $\beta^-$--decay of the
  neutron for three photon energy regions, calculated for the lifetime
  of the neutron $\tau_n = 879.6(1.1)\,{\rm s}$ \cite{Ivanov2013}. In
  the last column we give a relative contribution of the $1/M$
  corrections, caused by the weak magnetism and proton recoil. }
\end{table}

 Of course, the numerical values of the branching ratios should not be
 practically changed if we would use the world averaged value of the
 lifetime of the neutron $\tau_n = 880.2(1.0)\,{\rm s}$
 \cite{PDG2016}, which agrees perfectly well with the theoretical one
 $\tau_n = 879.6(1.1)\,{\rm s}$ \cite{Ivanov2013}.  From the
 comparison of the branching ratios of the neutron radiative
 $\beta^-$--decay, calculated to leading order in the large proton
 mass expansion, one may see that the contributions of the weak
 magnetism and proton recoil make up of about $0.70\,\%$, $0.66\,\%$
 and $0.60\,\%$ for the photon energy regions $15\,{\rm keV} \le
 \omega \le 340\,{\rm keV}$, $14\,{\rm keV} \le \omega \le 782\,{\rm
   keV}$ and $0.4\,{\rm keV} \le\omega \le 14\,{\rm keV}$,
 respectively.  Thus, at first glimpse, the contributions of the weak
 magnetism and proton recoil to the neutron radiative
 $\beta^-$--decays seem to be not very important. Moreover, that such
 contributions are small compared to the values of the statistic
 errors of the new experimental data $1.49\,\%$ and $3.95\,\%$ for the
 photon energy regions $14\,{\rm keV} \le \omega \le 782\,{\rm keV}$
 and $0.4\,{\rm keV} \le\omega \le 14\,{\rm keV}$, respectively. The
 account for the systematic errors, making up of about $4.78\,\%$ and
 $11.34\,\%$ of the new experimental values, measured for the photon
 energy regions $14\,{\rm keV} \le \omega \le 782\,{\rm keV}$ and
 $0.4\,{\rm keV} \le\omega \le 14\,{\rm keV}$, respectively, makes the
 contribution of the weak magnetism and proton recoil fully
 intangible. Nevertheless, we would like to emphasize that the values
 of the $1/M$ corrections, caused by the weak magnetism and proton
 recoil, to the branching ratios of the neutron radiative
 $\beta^-$--decay are large compared by a factor of 4 to the
 contribution of the weak magnetism and proton recoil of about
 $0.16\,\%$ to the neutron lifetime, calculated to order $1/M$ in
 \cite{Ivanov2013}. Thus, one may conclude that the weak magnetism and
 proton recoil, taken to order $1/M$, play more important role for the
 rate of the neutron radiative $\beta^-$--decay than for the rate of
 the neutron $\beta^-$--decay. On the whole they give a relative
 corrections of order $10^{-2}$.

Combing the statistic and systematic errors the experimental values
for the branching ratios we obtain ${\rm BR}^{(\exp)}_{\beta\gamma} =
3.35(16)\times 10^{-3}$ and ${\rm BR}^{(\exp)}_{\beta\gamma} =
5.82(66)\times 10^{-3}$ for the photon energy regions $14\,{\rm keV}
\le \omega \le 782\,{\rm keV}$ and $0.4\,{\rm keV} \le \omega \le
14\,{\rm keV}$, respectively. One may see that the theoretical values
of the branching ratios may agree with the experimental ones only
within 2 and 1.2 standard deviations, respectively. The theoretical
values of the branching ratios are below the experimental mean-values
by about $9\,\%$ and $13\,\%$. This leaves room for the analysis of
other contributions to the neutron radiative
$\beta^-$--decay. Following Bernard {\it et al.}  \cite{Bernard2004}
these contributions in the tree--approximation can be collected from
the baryon resonances \cite{PDG2016}. For example, the contribution of
the $\Delta(1232)$--resonance with spin and parity $J^{\pi} =
\frac{3}{2}^+$, the mass of which $m_{\Delta} \simeq 1232\,{\rm MeV}$
is not far from the proton mass \cite{PDG2016}, has been calculated by
Bernard {\it et al.}  \cite{Bernard2004}. According to Cooper {\it et
  al.}  \cite{Cooper2010}, the contribution of the
$\Delta(1232)$--resonance makes up only of about $0.5\,\%$ of the
branching ratio, measured for the photon energy region $15\,{\rm keV}
\le \omega \le 340\,{\rm keV}$ \cite{Cooper2010}. This implies that
other baryon resonances with heavier masses \cite{PDG2016} should give
contributions to the rate of the neutron radiative $\beta^-$--decay,
which are small compared even to that by the
$\Delta(1232)$--resonance.

Thus, one may expect that the theoretical analysis of the neutron
radiative $\beta^-$--decay, which can be performed in the Standard
Model and in the tree--approximation, may in principle change the rate
of the neutron radiative $\beta^-$--decay not stronger than by about
$1.5\,\%$ or even smaller. Hence, as a next step towards a precision
theoretical analysis of the neutron radiative $\beta^-$--decay, which
can be performed in the Standard Model, we may relate only to the
analysis beyond the tree--approximation. To our point of view this
should concern the radiative corrections to order $O(\alpha/\pi)$,
calculated to leading order in the large proton mass expansion
\cite{Ivanov2013}.  It is well--known \cite{Sirlin1967} that the
radiative corrections, calculated to order $O(\alpha/\pi)$ and to
leading order in the large proton mass expansion, change the rate of
the neutron $\beta^-$--decay by about $3.75\,\%$
\cite{Ivanov2013}. Taking into account that the corrections of the
weak magnetism and proton recoil of order $1/M$ to the neutron
radiative $\beta^-$--decay rate are by a factor 4 large compared to
the $1/M$ corrections to the rate of the neutron $\beta^-$--decay one
may expect that the relative contribution of the radiative corrections
of order $O(\alpha/\pi)$ to the rate of the neutron radiative
$\beta^-$--decay can be also substantially enhanced.  The first step
in the direction of the account for the radiative corrections of order
$O(\alpha/\pi)$ to the neutron radiative $\beta^-$--decay has been
made by Gardner and He \cite{Gardner2012,Gardner2013}.  However, the
results, obtained by Gardner and He \cite{Gardner2012,Gardner2013},
concern only the radiative corrections of order $O(\alpha/\pi)$ to
T--odd momentum correlations in the neutron radiative
$\beta^-$--decay.  We are planning to give a detailed analysis of the
radiative corrections of order $O(\alpha/\pi)$, allowing to describe
the rate of the neutron radiative $\beta^-$--decay to order
$O(\alpha^2/\pi^2)$, in our forthcoming publication.

\section{Acknowledgements}

The work of A. N. Ivanov was supported by the Austrian ``Fonds zur
F\"orderung der Wissenschaftlichen Forschung'' (FWF) under Contracts
I689-N16, I862-N20 and P26781-N20 and by the \"OAW within the New
Frontiers Groups Programme, NFP 2013/09. The work of R. H\"ollwieser
was supported by the Erwin Schr\"odinger Fellowship program of the
Austrian Science Fund FWF (``Fonds zur F\"orderung der
wissenschaftlichen Forschung'') under Contract No. J3425-N27. The work
of M. Wellenzohn was supported by the MA 23 (FH-Call 16) under the
project ``Photonik - Stiftungsprofessur f\"ur Lehre''.
 
\newpage

\section*{Appendix A: Amplitude of radiative $\beta^-$--decay 
of neutron, described by Feynman diagrams in Fig.\,1, with ``weak
magnetism'' and proton recoil corrections to order $1/M$}
\renewcommand{\theequation}{A-\arabic{equation}}
\setcounter{equation}{0}

The amplitude of the neutron radiative $\beta^-$--decay, given by
Eq.(\ref{eq:5}), we transcribe into the form
\cite{Ivanov2013,Ivanov2013b}
\begin{eqnarray}\label{eq:A.1}
\hspace{-0.3in}{\cal M}(n \to p e^- \bar{\nu}_e \gamma)_{\lambda'} &=&
\Big[\bar{u}_p(\vec{k}_p, \sigma_p) O_{\mu} u_n(\vec{k}_n,
  \sigma_n)\Big]
\Big[\bar{u}_e(\vec{k}_e,\sigma_e)\,\frac{1}{2k_e\cdot
    k}\,Q_e\,\gamma^{\mu} (1 - \gamma^5) v_{\nu}(\vec{k}_{\nu}, +
    \frac{1}{2})\Big]\nonumber\\
\hspace{-0.3in}&& - \Big[\bar{u}_p(\vec{k}_p, \sigma_p)\,Q_p
  \,\frac{1}{2k_p \cdot k}\,O_{\mu} u_n(\vec{k}_n,
  \sigma_n)\Big]\Big[\bar{u}_e(\vec{k}_e,\sigma_e) \gamma^{\mu} (1 -
  \gamma^5) v_{\nu}(\vec{k}_{\nu}, + \frac{1}{2})\Big],
\end{eqnarray}
where $\bar{u}_p(\vec{k}_p, \sigma_p)$, $u_n(\vec{k}_n, \sigma_n)$,
$\bar{u}_e(\vec{k}_e,\sigma_e)$ and $v_{\nu}(\vec{k}_{\nu}, +
\frac{1}{2})$ are the Dirac bispinor wave functions of the free
proton, neutron, electron and electron antineutrino with 3--momenta
$\vec{k}_p$, $\vec{k}_n = \vec{0}$, $\vec{k}_e$ and $\vec{k}_{\nu}$
and polarizations $\sigma_p = \pm 1$, $\sigma_n = \pm 1$, $\sigma_e =
\pm 1$ and $+ \frac{1}{2}$ \cite{Ivanov2013,Ivanov2014}, respectively,
$Q_e$ and $Q_p$ are defined by
\begin{eqnarray}\label{eq:A.2}
\hspace{-0.3in}Q_e &=& 2 \varepsilon^{*}_{\lambda'}\cdot k_e +
\hat{\varepsilon}^*_{\lambda'}\hat{k},\nonumber\\
\hspace{-0.3in}Q_p &=& 2 \varepsilon^{*}_{\lambda'}\cdot k_p +
\hat{\varepsilon}^*_{\lambda'}\hat{k}.
\end{eqnarray}
For the derivation of Eq.(\ref{eq:A.1}) we have used the Dirac
equations for the free proton and electron. Replacing
$\varepsilon^{*}_{\lambda'} \to k$ and using $k^2 = 0$ we get ${\cal
  M}(n \to p e^- \bar{\nu}_e \gamma)|_{\varepsilon^{*}_{\lambda'} \to
  k} = 0$. The matrix $Q_{\mu}$ is given by Eq.(\ref{eq:6}). For the
analysis of the neutron radiative $\beta^-$--decay in the
non--relativistic approximation we have to rewrite the amplitude
Eq.(\ref{eq:A.1}) in terms of time and space components of the matrix
$O_{\mu} = (O_0, -\vec{O}\,)$ \cite{Ivanov2013}. We get
\begin{eqnarray}\label{eq:A.3}
\hspace{-0.3in}{\cal M}(n \to p e^- \bar{\nu}_e \gamma)_{\lambda'} &=&
\Big[\bar{u}_p(\vec{k}_p, \sigma_p) O_0 u_n(\vec{k}_n,
  \sigma_n)\Big]
\Big[\bar{u}_e(\vec{k}_e,\sigma_e)\,\frac{1}{2k_e\cdot
    k}\,Q_e\,\gamma^0 (1 - \gamma^5) v_{\nu}(\vec{k}_{\nu}, +
  \frac{1}{2})\Big]\nonumber\\
\hspace{-0.3in}&& - \Big[\bar{u}_p(\vec{k}_p, \sigma_p)\,\vec{O}\,
  u_n(\vec{k}_n, \sigma_n)\Big] \cdot
\Big[\bar{u}_e(\vec{k}_e,\sigma_e)\,\frac{1}{2k_e\cdot
    k}\,Q_e\,\vec{\gamma}\,(1 - \gamma^5) v_{\nu}(\vec{k}_{\nu}, +
  \frac{1}{2})\Big]\nonumber\\
\hspace{-0.3in}&& - \Big[\bar{u}_p(\vec{k}_p,
  \sigma_p) Q_p\frac{1}{2k_p\cdot k}\, O_0 u_n(\vec{k}_n,
  \sigma_n)\Big] \Big[\bar{u}_e(\vec{k}_e,\sigma_e)\,\gamma^0 (1 -
  \gamma^5) v_{\nu}(\vec{k}_{\nu}, + \frac{1}{2})\Big]\nonumber\\
\hspace{-0.3in}&& + \Big[\bar{u}_p(\vec{k}_p,
  \sigma_p)\,\frac{1}{2k_p\cdot k}\,\vec{O}\, u_n(\vec{k}_n,
  \sigma_n)\Big] \cdot
\Big[\bar{u}_e(\vec{k}_e,\sigma_e)\,\vec{\gamma}\,(1 - \gamma^5)
  v_{\nu}(\vec{k}_{\nu}, + \frac{1}{2})\Big].
\end{eqnarray}
The time $O_0$ and spatial $\vec{O}$ components of the matrix
$O_{\mu}$ we determine to order $1/M$ in the large $M$ expansion
\cite{Ivanov2013}
\begin{eqnarray}\label{eq:A.4}
 \hspace{-0.3in} O_0 =\left(\begin{array}{ccc} 1 & {\displaystyle
     \lambda + \frac{\kappa}{2 M}\,(\vec{\sigma}\cdot
     \vec{k}_p)}\\ {\displaystyle - \lambda + \frac{\kappa}{2
       M}\,(\vec{\sigma}\cdot \vec{k}_p) } & - 1\\
    \end{array}\right)
\end{eqnarray}
and 
\begin{eqnarray}\label{eq:A.5}
 \hspace{-0.3in}&& \vec{O} =\left(\begin{array}{ccc} {\displaystyle
     \lambda \vec{\sigma} + i\,\frac{\kappa}{2 M}\,(\vec{\sigma}\times
     \vec{k}_p)} & {\displaystyle \vec{\sigma}\,\Big(1 -
     \frac{\kappa}{2 M}\,(E_e + E_{\nu} +
     \omega)\Big)}\\ {\displaystyle -\,\vec{\sigma}\,\Big(1 +
     \frac{\kappa}{2 M}\,(E_e + E_{\nu} + \omega)\Big)} & {\displaystyle -
     \lambda \vec{\sigma}+ i \frac{\kappa}{2 M}\,(\vec{\sigma}\times
     \vec{k}_p)} \\
    \end{array}\right).
\end{eqnarray}
For the derivation of Eq.(\ref{eq:A.4}) and Eq.(\ref{eq:5}) we have
kept only the terms of order $1/M$ and used the energy conservation
$m_n = E_p + E_e + E_{\nu} + \omega$, where $m_n$, $E_p =
\sqrt{m^2_p + \vec{k}^{\,2}_p}$, $E_e$, $E_{\nu}$ and $\omega$
are the neutron mass and the proton, electron, antineutrino and photon
energies, respectively. Then, $\vec{\sigma}$ are the Pauli $2\times 2$
matrices \cite{IZ1980}.

 For the calculation of the amplitude of the neutron radiative
 $\beta^-$--decay we define the Dirac bispinor wave functions of the
 neutron and the proton as follows
\begin{eqnarray}\label{labelA.6}
 \hspace{-0.3in} u_n(\vec{0},\sigma_n) = \sqrt{2
   m_n}\Big(\begin{array}{c}\varphi_n \\ 0
 \end{array}\Big) \quad,\quad u_p(\vec{k}_p,\sigma_p) = \sqrt{E_p +
 m_p}\left(\begin{array}{c}\varphi_p \\ {\displaystyle
 \frac{\vec{\sigma}\cdot \vec{k}_p}{E_p + m_p}\,\varphi_p }
 \end{array}\right),
\end{eqnarray}
where the Pauli spinorial wave functions $\varphi_n$ and $\varphi_p$
depend on the polarizations $\sigma_n$ and $\sigma_p$,
respectively. The matrix elements $[\bar{u}_p(\vec{k}_p,\sigma_p) O_0
  u_n(\vec{0},\sigma_n)]$ and $[\bar{u}_p(\vec{k}_p, \sigma_p) \vec{O}
  u_n(\vec{0},\sigma_n)]$ are equal to \cite{Ivanov2013}
\begin{eqnarray}\label{labelA.7}
 \hspace{-0.3in}&&[\bar{u}_p(\vec{k}_p, \sigma_p) O_0
   u_n(\vec{0},\sigma_n)] = \sqrt{2 m_n(E_p +
   m_p)}\,\Big\{[\varphi^{\dagger}_p\varphi_n] + \frac{\lambda}{2
   M}\,[\varphi^{\dagger}_p(\vec{\sigma}\cdot
   \vec{k}_p)\varphi_n]\Big\}
\end{eqnarray}
and 
\begin{eqnarray}\label{labelA.8}
 \hspace{-0.3in}&&[\bar{u}_p (\vec{k}_p, \sigma_p)\vec{O}
   u_n(\vec{0},\sigma_n)] = \sqrt{2 m_n(E_p + m_p)}\,\Big\{\lambda
        [\varphi^{\dagger}_p\vec{\sigma}\,\varphi_n] + i\,\frac{\kappa
        }{2 M}\,[\varphi^{\dagger}_p(\vec{\sigma}\times
          \vec{k}_p)\varphi_n] + \frac{1}{2
          M}\,[\varphi^{\dagger}_p(\vec{\sigma}\cdot
          \vec{k}_p)\vec{\sigma}\,\varphi_n]\Big\}.
\end{eqnarray}
For the calculation of the matrix elements
\begin{eqnarray}\label{labelA.9}
 \hspace{-0.3in}[\bar{u}_p(\vec{k}_p, \sigma_p)Q_p\frac{1}{2k_p\cdot
     k} O_0 u_n(\vec{0},\sigma_n)]\;,\; [\bar{u}_p (\vec{k}_p,
   \sigma_p)Q_p\frac{1}{2k_p\cdot k} \vec{O} u_n(\vec{0},\sigma_n)]
\end{eqnarray}
we have to define the products $Q_p (1/2k_p\cdot k )\, Q_0$ and
$Q_p (1/2k_p\cdot k)\, \vec{O}$. For this aim we define the matrix
$Q_p (1/2k_p\cdot k)$ as follows
\begin{eqnarray}\label{labelA.10}
 \hspace{-0.3in} O_p\,\frac{1}{2 k_p\cdot k}
 =\left(\begin{array}{ccc}{\displaystyle
     \frac{2\varepsilon^*_{\lambda'}\cdot k_p}{2k_p\cdot k} +
     i\,\frac{\vec{\sigma}\cdot
       (\vec{\varepsilon}^{\,*}_{\lambda'}\times \vec{k}\,)}{2k_p\cdot
       k}} & {\displaystyle \frac{-
       \varepsilon^{0*}_{\lambda'}(\vec{\sigma}\cdot \vec{k}\,) +
       \omega (\vec{\varepsilon}^{\,*}_{\lambda'}\cdot
       \vec{\sigma}\,)}{2k_p\cdot k}}\\ {\displaystyle \frac{-
       \varepsilon^{0*}_{\lambda'}(\vec{\sigma}\cdot \vec{k}\,) +
       \omega (\vec{\varepsilon}^{\,*}_{\lambda'}\cdot
       \vec{\sigma}\,)}{2k_p\cdot k}} & {\displaystyle
     \frac{2\varepsilon^*_{\lambda'}\cdot k_p}{2k_p\cdot k} +
     i\,\frac{\vec{\sigma}\cdot
       (\vec{\varepsilon}^{\,*}_{\lambda'}\times \vec{k}\,)}{2k_p\cdot
       k}}\\
    \end{array}\right),
\end{eqnarray}
where we have denoted $\varepsilon^{*\alpha}_{\lambda'} =
(\varepsilon^{0*}_{\lambda'}, \vec{\varepsilon}^{\,*}_{\lambda'})$ and
$k^{\alpha} = (\omega, \vec{k}\,)$. Replacing
$\varepsilon^{*\alpha}_{\lambda'} = (\varepsilon^{0*}_{\lambda'},
\vec{\varepsilon}^{\,*}_{\lambda'}) \to k^{\alpha} = (\omega,
\vec{k}\,)$ we get the matrix $Q_p (1/2k_p\cdot k)$ equal to a unit
matrix $Q_p (1/2k_p\cdot k)|_{\varepsilon^{*\alpha}_{\lambda'} \to k}
= 1$. Thus, the product $Q_p\,(1/2k_p\cdot k)\,Q_0$ is given by
\begin{eqnarray}\label{labelA.11}
 \hspace{-0.3in} O_p \frac{1}{2 k_p\cdot k}Q_0 =
 \left(\begin{array}{ccc}{\displaystyle \Big(O_p \frac{1}{2 k_p\cdot
       k}Q_0\Big)_{11}} & {\displaystyle \Big(O_p \frac{1}{2 k_p\cdot
       k}Q_0\Big)_{12}}\\ {\displaystyle \Big(O_p \frac{1}{2 k_p\cdot
       k}Q_0\Big)_{21} } & {\displaystyle \Big(O_p \frac{1}{2 k_p\cdot
       k}Q_0\Big)_{22}}\\
    \end{array}\right),
\end{eqnarray}
where we have denoted
\begin{eqnarray}\label{labelA.12}
\Big(O_p \frac{1}{2 k_p\cdot k}Q_0\Big)_{11} &=&
\frac{\varepsilon^{0*}_{\lambda'}}{\omega} +
i\,\frac{\vec{\sigma}\cdot (\vec{\varepsilon}^{\,*}_{\lambda'}\times
  \vec{k}\,)}{2 M \omega} + \lambda \frac{
  \varepsilon^{0*}_{\lambda'}(\vec{\sigma}\cdot \vec{k}\,) - \omega
  (\vec{\varepsilon}^{\,*}_{\lambda'}\cdot \vec{\sigma}\,)}{2 M
  \omega},\nonumber\\ \Big(O_p \frac{1}{2 k_p\cdot k}Q_0\Big)_{12} &=&
\frac{\varepsilon^{0*}_{\lambda'}}{\omega}\Big(\lambda +
\frac{\kappa}{2 M}\,(\vec{\sigma}\cdot \vec{k}_p)\Big) +
i\,\lambda\,\frac{\vec{\sigma}\cdot
  (\vec{\varepsilon}^{\,*}_{\lambda'}\times \vec{k}\,)}{2 M \omega}+
\frac{ \varepsilon^{0*}_{\lambda'}(\vec{\sigma}\cdot \vec{k}\,) -
  \omega (\vec{\varepsilon}^{\,*}_{\lambda'}\cdot \vec{\sigma}\,)}{2 M
  \omega},\nonumber\\ \Big(O_p \frac{1}{2 k_p\cdot k}Q_0\Big)_{21} &=&
\frac{\varepsilon^{0*}_{\lambda'}}{\omega}\Big(- \lambda +
\frac{\kappa}{2 M}\,(\vec{\sigma}\cdot \vec{k}_p)\Big) -
i\,\lambda\,\frac{\vec{\sigma}\cdot
  (\vec{\varepsilon}^{\,*}_{\lambda'}\times \vec{k}\,)}{2 M \omega} -
\frac{ \varepsilon^{0*}_{\lambda'}(\vec{\sigma}\cdot \vec{k}\,) -
  \omega (\vec{\varepsilon}^{\,*}_{\lambda'}\cdot \vec{\sigma}\,)}{2 M
  \omega},\nonumber\\ \Big(O_p \frac{1}{2 k_p\cdot k}Q_0\Big)_{22} &=&
- \frac{\varepsilon^{0*}_{\lambda'}}{\omega} -
i\,\frac{\vec{\sigma}\cdot (\vec{\varepsilon}^{\,*}_{\lambda'}\times
  \vec{k}\,)}{2 M \omega} - \lambda \frac{
  \varepsilon^{0*}_{\lambda'}(\vec{\sigma}\cdot \vec{k}\,) - \omega
  (\vec{\varepsilon}^{\,*}_{\lambda'}\cdot \vec{\sigma}\,)}{2 M
  \omega}.
\end{eqnarray}
In turn, the product $Q_p\,(1/2k_p\cdot k)\,\vec{O}$ reads
\begin{eqnarray}\label{labelA.13}
 \hspace{-0.3in} O_p \frac{1}{2 k_p\cdot k}\vec{Q} =
 \left(\begin{array}{ccc}{\displaystyle \Big(O_p \frac{1}{2 k_p\cdot
       k} \vec{Q}\,\Big)_{11}} & {\displaystyle \Big(O_p \frac{1}{2
       k_p\cdot k} \vec{Q}\,\Big)_{12}}\\ {\displaystyle \Big(O_p
     \frac{1}{2 k_p\cdot k} \vec{Q}\,\Big)_{21} } & {\displaystyle
     \Big(O_p \frac{1}{2 k_p\cdot k} \vec{Q}\,\Big)_{22}}\\
    \end{array}\right),
\end{eqnarray}
where we have denoted
\begin{eqnarray}\label{labelA.14}
\Big(O_p \frac{1}{2 k_p\cdot k} \vec{Q}\,\Big)_{11} &=&
\frac{\varepsilon^{0*}_{\lambda'}}{\omega}\Big(\lambda \vec{\sigma} +
i\,\frac{\kappa}{2 M}\,(\vec{\sigma}\times \vec{k}_p)\Big) +
i\,\lambda\,\frac{\vec{\sigma}\cdot
  (\vec{\varepsilon}^{\,*}_{\lambda'}\times \vec{k}\,)}{2 M
  \omega}\,\vec{\sigma} + \lambda \,\frac{
  \varepsilon^{0*}_{\lambda'}(\vec{\sigma}\cdot \vec{k}\,) - \omega
  (\vec{\varepsilon}^{\,*}_{\lambda'}\cdot \vec{\sigma}\,)}{2 M
  \omega}\,\vec{\sigma},\nonumber\\ \Big(O_p \frac{1}{2 k_p\cdot k}
\vec{Q}\,\Big)_{12} &=&
\frac{\varepsilon^{0*}_{\lambda'}}{\omega}\Big(1 - \frac{\kappa}{2
  M}\,(E_e + E + \omega)\Big) + i\,\frac{\vec{\sigma}\cdot
  (\vec{\varepsilon}^{\,*}_{\lambda'}\times \vec{k}\,)}{2 M
  \omega}\,\vec{\sigma} + \frac{
  \varepsilon^{0*}_{\lambda'}(\vec{\sigma}\cdot \vec{k}\,) - \omega
  (\vec{\varepsilon}^{\,*}_{\lambda'}\cdot \vec{\sigma}\,)}{2 M
  \omega}\,\vec{\sigma},\nonumber\\ \Big(O_p \frac{1}{2 k_p\cdot k}
\vec{Q}\,\Big)_{21} &=& -
\frac{\varepsilon^{0*}_{\lambda'}}{\omega}\Big(1 + \frac{\kappa}{2
  M}\,(E_e + E + \omega)\Big) - i\,\frac{\vec{\sigma}\cdot
  (\vec{\varepsilon}^{\,*}_{\lambda'}\times \vec{k}\,)}{2 M
  \omega}\,\vec{\sigma} - \lambda\,\frac{
  \varepsilon^{0*}_{\lambda'}(\vec{\sigma}\cdot \vec{k}\,) - \omega
  (\vec{\varepsilon}^{\,*}_{\lambda'}\cdot \vec{\sigma}\,)}{2 M
  \omega}\,\vec{\sigma},\nonumber\\ \Big(O_p \frac{1}{2 k_p\cdot k}
\vec{Q}\,\Big)_{22} &=& -
\frac{\varepsilon^{0*}_{\lambda'}}{\omega}\Big(\lambda \vec{\sigma} +
i\,\frac{\kappa}{2 M}\,(\vec{\sigma}\times \vec{k}_p)\Big) -
i\,\lambda\,\frac{\vec{\sigma}\cdot
  (\vec{\varepsilon}^{\,*}_{\lambda'}\times \vec{k}\,)}{2 M
  \omega}\,\vec{\sigma} - \frac{
  \varepsilon^{0*}_{\lambda'}(\vec{\sigma}\cdot \vec{k}\,) - \omega
  (\vec{\varepsilon}^{\,*}_{\lambda'}\cdot \vec{\sigma}\,)}{2 M
  \omega}\,\vec{\sigma}.
\end{eqnarray}
We would like to remind that because of the relation
$\varepsilon^{*}_{\lambda'}\cdot k = 0$ the time--component of the
polarization vector is equal to $\varepsilon^{0*}_{\lambda'} =
(\vec{\varepsilon}^{\,*}_{\lambda'}\cdot \vec{k}\,)/\omega$. Replacing
$\vec{\varepsilon}^{\,*}_{\lambda'} \to \vec{k}$ and using
$|\vec{k}\,| = \omega$ we get $Q_p\,(1/2k_p\cdot
k)\,Q_0|_{\vec{\varepsilon}^{\,*}_{\lambda'} \to \vec{k}} = Q_0$ and
$Q_p\,(1/2k_p\cdot k)\,\vec{O}\,|_{\vec{\varepsilon}^{\,*}_{\lambda'}
  \to \vec{k}} = \vec{O}$, respectively. Now we may define the matrix
elements
\begin{eqnarray}\label{labelA.15}
 \hspace{-0.3in}&&[\bar{u}_p(\vec{k}_p, \sigma_p)Q_p\frac{1}{2k_p\cdot
     k} O_0 u_n(\vec{0},\sigma_n)] = \sqrt{2 m_n(E_p +
   m_p)}\,\Big\{\varphi^{\dagger}_p\Big(O_p \frac{1}{2 k_p\cdot k}
 O_0\Big)_{11}\varphi_n +
 \lambda\,\frac{\varepsilon^{0*}_{\lambda'}}{2M\omega}\,
 \varphi^{\dagger}_p(\vec{\sigma} \cdot \vec{k}_p)
 \varphi_n\Big\}=\nonumber\\
\hspace{-0.3in}&&= \sqrt{2 m_n(E_p +
  m_p)}\,\Big[\varphi^{\dagger}_p\Big(\frac{\varepsilon^{0*}_{\lambda'}}{\omega}
  + i\,\frac{\vec{\sigma}\cdot
    (\vec{\varepsilon}^{\,*}_{\lambda'}\times \vec{k}\,)}{2 M \omega}
  + \lambda \frac{ \varepsilon^{0*}_{\lambda'}(\vec{\sigma}\cdot
    \vec{k}\,) - \omega (\vec{\varepsilon}^{\,*}_{\lambda'}\cdot
    \vec{\sigma}\,)}{2 M \omega} +
  \lambda\,\frac{\varepsilon^{0*}_{\lambda'}}{2M\omega}\,(\vec{\sigma}\cdot
  \vec{k}_p)\Big)\varphi_n\Big]
\end{eqnarray}
and 
\begin{eqnarray}\label{labelA.16}
 \hspace{-0.3in}&&[\bar{u}_p (\vec{k}_p,
   \sigma_p)Q_p\frac{1}{2k_p\cdot k} \vec{O} u_n(\vec{0},\sigma_n)] =
 \sqrt{2 m_n(E_p + m_p)}\,\Big\{\varphi^{\dagger}_p\Big(O_p \frac{1}{2
   k_p\cdot k} \vec{Q}\,\Big)_{11}\varphi_n +
 \lambda\,\frac{\varepsilon^{0*}_{\lambda'}}{2M\omega}\,
 \varphi^{\dagger}_p(\vec{\sigma} \cdot \vec{k}_p)
 \varphi_n\Big\}=\nonumber\\
\hspace{-0.3in}&&= \sqrt{2 m_n(E_p + m_p)}\,\Big[\varphi^{\dagger}_p
  \Big(\frac{\varepsilon^{0*}_{\lambda'}}{\omega}\Big(\lambda
  \vec{\sigma} + i\,\frac{\kappa}{2 M}\,(\vec{\sigma}\times
  \vec{k}_p)\Big) + i\,\lambda\,\frac{\vec{\sigma}\cdot
    (\vec{\varepsilon}^{\,*}_{\lambda'}\times \vec{k}\,)}{2 M
    \omega}\,\vec{\sigma} + \lambda \,\frac{
    \varepsilon^{0*}_{\lambda'}(\vec{\sigma}\cdot \vec{k}\,) - \omega
    (\vec{\varepsilon}^{\,*}_{\lambda'}\cdot \vec{\sigma}\,)}{2 M
    \omega}\,\vec{\sigma}\nonumber\\
\hspace{-0.3in}&& +
\lambda\,\frac{\varepsilon^{0*}_{\lambda'}}{2M\omega}\,(\vec{\sigma}\cdot
\vec{k}_p)\,\vec{\sigma}\,\Big)\varphi_n\Big].
\end{eqnarray}
The amplitude of the neutron radiative $\beta^-$--decay, calculated to
order $1/M$, caused by the weak magnetism and proton recoil, is equal
to
\begin{eqnarray}\label{eq:A.17}
\hspace{-0.3in}&&{\cal M}(n \to p e^- \bar{\nu}_e \gamma)_{\lambda'} = \sqrt{2
  m_n(E_p + m_p)}\,\Big\{\Big[\varphi^{\dagger}_p\Big(1 +
  \frac{\lambda}{2 M}\,(\vec{\sigma}\cdot
  \vec{k}_p)\Big)\varphi_n\Big]
\Big[\bar{u}_e(\vec{k}_e,\sigma_e)\,\frac{1}{2k_e\cdot
    k}\,Q_e\,\gamma^0 (1 - \gamma^5) v_{\nu}(\vec{k}, +
  \frac{1}{2})\Big]\nonumber\\
\hspace{-0.3in}&& - \Big[\varphi^{\dagger}_p\Big(\lambda\,\vec{\sigma}
  + i\,\frac{\kappa }{2 M}\,(\vec{\sigma}\times \vec{k}_p) +
  \frac{1}{2 M}\,(\vec{\sigma}\cdot \vec{k}_p)\vec{\sigma}\,\Big)
  \varphi_n\Big] \cdot
\Big[\bar{u}_e(\vec{k}_e,\sigma_e)\,\frac{1}{2k_e\cdot
    k}\,Q_e\,\vec{\gamma}\,(1 - \gamma^5) v_{\nu}(\vec{k}, +
  \frac{1}{2})\Big]\nonumber\\
\hspace{-0.3in}&& -
\Big[\varphi^{\dagger}_p\Big(\frac{\varepsilon^{0*}_{\lambda'}}{\omega}
  + i\,\frac{\vec{\sigma}\cdot
    (\vec{\varepsilon}^{\,*}_{\lambda'}\times \vec{k}\,)}{2 M \omega}
  + \lambda \frac{ \varepsilon^{0*}_{\lambda'}(\vec{\sigma}\cdot
    \vec{k}\,) - \omega (\vec{\varepsilon}^{\,*}_{\lambda'}\cdot
    \vec{\sigma}\,)}{2 M \omega} +
  \lambda\,\frac{\varepsilon^{0*}_{\lambda'}}{2M\omega}\,(\vec{\sigma}\cdot
  \vec{k}_p)\Big)\varphi_n\Big] 
\Big[\bar{u}_e(\vec{k}_e,\sigma_e)\,\gamma^0 (1 - \gamma^5)
  v_{\nu}(\vec{k}, + \frac{1}{2})\Big]\nonumber\\
\hspace{-0.3in}&& + \Big[\varphi^{\dagger}_p
  \Big(\frac{\varepsilon^{0*}_{\lambda'}}{\omega}\Big(\lambda
  \vec{\sigma} + i\,\frac{\kappa}{2 M}\,(\vec{\sigma}\times
  \vec{k}_p)\Big) + i\,\lambda\,\frac{\vec{\sigma}\cdot
    (\vec{\varepsilon}^{\,*}_{\lambda'}\times \vec{k}\,)}{2 M
    \omega}\,\vec{\sigma} + \lambda \,\frac{
    \varepsilon^{0*}_{\lambda'}(\vec{\sigma}\cdot \vec{k}\,) - \omega
    (\vec{\varepsilon}^{\,*}_{\lambda'}\cdot \vec{\sigma}\,)}{2 M
    \omega}\,\vec{\sigma} +
  \lambda\,\frac{\varepsilon^{0*}_{\lambda'}}{2M\omega}\,(\vec{\sigma}\cdot
  \vec{k}_p)\,\vec{\sigma}\,\Big)\varphi_n\Big]\nonumber\\
\hspace{-0.3in}&&\cdot
\Big[\bar{u}_e(\vec{k}_e,\sigma_e)\,\vec{\gamma}\,(1 - \gamma^5)
  v_{\nu}(\vec{k}, + \frac{1}{2})\Big]\Big\}.
\end{eqnarray}
Since the factor $\sqrt{2m_m(E_p + m_p)}$ to order $1/M$ is equal to
\begin{eqnarray}\label{eq:A.18}
\hspace{-0.3in}\sqrt{2m_m(E_p + m_p)} = 2m_n\Big(1 - \frac{E_e +
  E_{\nu} + \omega}{2M}\Big),
\end{eqnarray}
the amplitude Eq.(\ref{eq:A.17}) can be transcribed into the form
\begin{eqnarray}\label{eq:A.19}
\hspace{-0.3in}&&{\cal M}(n \to p e^- \bar{\nu}_e \gamma)_{\lambda'} =
2m_n\,\Big\{\Big[\varphi^{\dagger}_p\Big(1 - \frac{E_e + E_{\nu} +
    \omega}{2M} + \frac{\lambda}{2 M}\,(\vec{\sigma}\cdot
  \vec{k}_p)\Big)\varphi_n\Big]
\Big[\bar{u}_e(\vec{k}_e,\sigma_e)\,\frac{1}{2k_e\cdot
    k}\,Q_e\,\gamma^0 (1 - \gamma^5) v_{\nu}(\vec{k}, +
  \frac{1}{2})\Big]\nonumber\\
\hspace{-0.3in}&& - \Big[\varphi^{\dagger}_p\Big(\lambda\,\vec{\sigma}
  - \lambda\,\frac{E_e + E_{\nu} + \omega}{2M}\,\vec{\sigma} +
  i\,\frac{\kappa }{2 M}\,(\vec{\sigma}\times \vec{k}_p) + \frac{1}{2
    M}\,(\vec{\sigma}\cdot \vec{k}_p)\vec{\sigma}\,\Big)
  \varphi_n\Big] \cdot
\Big[\bar{u}_e(\vec{k}_e,\sigma_e)\,\frac{1}{2k_e\cdot
    k}\,Q_e\,\vec{\gamma}\,(1 - \gamma^5) v_{\nu}(\vec{k}, +
  \frac{1}{2})\Big]\nonumber\\
\hspace{-0.3in}&& -
\Big[\varphi^{\dagger}_p\Big(\frac{\varepsilon^{0*}_{\lambda'}}{\omega}
  - \frac{\varepsilon^{0*}_{\lambda'}}{\omega}\frac{E_e +
    E_{\nu} + \omega}{2M} + i\,\frac{\vec{\sigma}\cdot
    (\vec{\varepsilon}^{\,*}_{\lambda'}\times \vec{k}\,)}{2 M \omega}
  + \lambda \frac{ \varepsilon^{0*}_{\lambda'}(\vec{\sigma}\cdot
    \vec{k}\,) - \omega (\vec{\varepsilon}^{\,*}_{\lambda'}\cdot
    \vec{\sigma}\,)}{2 M \omega} +
  \lambda\,\frac{\varepsilon^{0*}_{\lambda'}}{2M\omega}\,(\vec{\sigma}\cdot
  \vec{k}_p)\Big)\varphi_n\Big]\nonumber\\
\hspace{-0.3in}&&\times\, \Big[\bar{u}_e(\vec{k}_e,\sigma_e)\,\gamma^0
  (1 - \gamma^5) v_{\nu}(\vec{k}, + \frac{1}{2})\Big] +
\Big[\varphi^{\dagger}_p
  \Big(\frac{\varepsilon^{0*}_{\lambda'}}{\omega}\Big(\lambda\,
  \vec{\sigma} - \lambda\,\frac{E_e + E_{\nu} + \omega}{2M}\, \vec{\sigma} +
  i\,\frac{\kappa}{2 M}\,(\vec{\sigma}\times \vec{k}_p)\Big) +
  i\,\lambda\,\frac{\vec{\sigma}\cdot
    (\vec{\varepsilon}^{\,*}_{\lambda'}\times \vec{k}\,)}{2 M
    \omega}\,\vec{\sigma}\nonumber\\
\hspace{-0.3in}&& + \lambda \,\frac{
  \varepsilon^{0*}_{\lambda'}(\vec{\sigma}\cdot \vec{k}\,) - \omega
  (\vec{\varepsilon}^{\,*}_{\lambda'}\cdot \vec{\sigma}\,)}{2 M
  \omega}\,\vec{\sigma} +
\lambda\,\frac{\varepsilon^{0*}_{\lambda'}}{2M\omega}\,(\vec{\sigma}\cdot
\vec{k}_p)\,\vec{\sigma}\,\Big)\varphi_n\Big] \cdot
\Big[\bar{u}_e(\vec{k}_e,\sigma_e)\,\vec{\gamma}\,(1 - \gamma^5)
  v_{\nu}(\vec{k}, + \frac{1}{2})\Big]\Big\}.
\end{eqnarray}
The hermitian conjugate amplitude is equal to
\begin{eqnarray*}
\hspace{-0.3in}&&{\cal M}^{\dagger}(n \to p e^- \bar{\nu}_e
\gamma)_{\lambda'} = \nonumber\\
\hspace{-0.3in}&& = 2m_n\,\Big\{\Big[\varphi^{\dagger}_n\Big(1 -
  \frac{E_e + E_{\nu} + \omega}{2M} + \frac{\lambda}{2
    M}\,(\vec{\sigma}\cdot \vec{k}_p)\Big)\varphi_p\Big]
\Big[\bar{v}_{\nu}(\vec{k}, + \frac{1}{2})\,\frac{1}{2k_e\cdot
    k}\,\gamma^0\, \bar{Q}_e (1 - \gamma^5)
  u_e(\vec{k}_e,\sigma_e)\Big]\nonumber\\
\hspace{-0.3in}&& - \Big[\varphi^{\dagger}_n\Big(\lambda\,\vec{\sigma}
  - \lambda\,\frac{E_e + E_{\nu} + \omega}{2M}\,\vec{\sigma} -
  i\,\frac{\kappa }{2 M}\,(\vec{\sigma}\times \vec{k}_p) + \frac{1}{2
    M}\,\vec{\sigma}\,(\vec{\sigma}\cdot \vec{k}_p)\Big)
  \varphi_p\Big] \cdot \Big[\bar{v}_{\nu}(\vec{k}, +
  \frac{1}{2})\,\frac{1}{2k_e\cdot k}\,\vec{\gamma}\,\bar{Q}_e\,(1 -
  \gamma^5) u_e(\vec{k}_e,\sigma_e)\Big]\nonumber\\
\hspace{-0.3in}&& -
\Big[\varphi^{\dagger}_n\Big(\frac{\varepsilon^{0}_{\lambda'}}{\omega}
  - \frac{\varepsilon^{0}_{\lambda'}}{\omega}\frac{E_e + E_{\nu} +
    \omega}{2M} - i\,\frac{\vec{\sigma}\cdot
    (\vec{\varepsilon}_{\lambda'}\times \vec{k}\,)}{2 M \omega}
  + \lambda \frac{ \varepsilon^{0}_{\lambda'}(\vec{\sigma}\cdot
    \vec{k}\,) - \omega (\vec{\varepsilon}_{\lambda'}\cdot
    \vec{\sigma}\,)}{2 M \omega} +
  \lambda\,\frac{\varepsilon^{0}_{\lambda'}}{2 M\omega}\,(\vec{\sigma}\cdot
  \vec{k}_p)\Big)\varphi_p\Big]\nonumber\\
\end{eqnarray*}
\begin{eqnarray}\label{eq:A.20}
\hspace{-0.3in}&&\times\, \Big[\bar{v}_{\nu}(\vec{k}, +
  \frac{1}{2})\,\gamma^0 (1 - \gamma^5) u_e(\vec{k}_e,\sigma_e)\Big] +
\Big[\varphi^{\dagger}_n
  \Big(\frac{\varepsilon^{0}_{\lambda'}}{\omega}\Big(\lambda\,
  \vec{\sigma} - \lambda\,\frac{E_e + E_{\nu} + \omega}{2M}\,
  \vec{\sigma} - i\,\frac{\kappa}{2 M}\,(\vec{\sigma}\times
  \vec{k}_p)\Big) - i\,\lambda\,\vec{\sigma}\,\frac{\vec{\sigma}\cdot
    (\vec{\varepsilon}_{\lambda'}\times \vec{k}\,)}{2 M
    \omega}\nonumber\\
\hspace{-0.3in}&& + \lambda \,\vec{\sigma}\,\frac{
  \varepsilon^{0}_{\lambda'}(\vec{\sigma}\cdot \vec{k}\,) - \omega
  (\vec{\varepsilon}_{\lambda'}\cdot \vec{\sigma}\,)}{2 M \omega} +
\lambda\,\frac{\varepsilon^{0}_{\lambda'}}{2M\omega} \,\vec{\sigma}
\,(\vec{\sigma}\cdot \vec{k}_p)\,\Big)\varphi_p\Big] \cdot
\Big[\bar{v}_{\nu}(\vec{k}, + \frac{1}{2})\,\vec{\gamma}\,(1 -
  \gamma^5) u_e(\vec{k}_e,\sigma_e)\Big]\Big\},
\end{eqnarray}
where $\bar{Q}_e = 2 k_e\cdot k + \hat{k}\hat{\varepsilon}_{\lambda'}$
\cite{Ivanov2013}. The squared absolute value of the amplitude
Eq.(\ref{eq:A.19}) summed over the polarizations of the proton and
electron and averaged over polarizations of the neutron is equal to
\begin{eqnarray}\label{eq:A.21}
\hspace{-0.3in}&&\sum_{\rm pol}\frac{|{\cal M}(n \to p e^- \bar{\nu}_e
  \gamma)_{\lambda'}|^2}{4m^2_n} = \frac{1}{(k_e\cdot k)^2}\,\Big(1 -
\frac{E_e + E_{\nu} + \omega}{M}\Big)\,{\rm tr}\{(\hat{k}_e +
m_e)\,Q_e \gamma^0\hat{k}_{\nu}\gamma^0 \bar{Q}_e (1 -
\gamma^5)\} - \frac{1}{(k_e\cdot k)^2}\,\frac{\lambda^2 +
  1}{2M}\nonumber\\
\hspace{-0.3in}&&\times\,\vec{k}_p\cdot{\rm tr}\{(\hat{k}_e +
m_e)\,Q_e \gamma^0\hat{k}_{\nu}\vec{\gamma}\, \bar{Q}_e (1 -
\gamma^5)\} -
2\,\frac{\varepsilon^{0}_{\lambda'}}{\omega}\,\frac{1}{k_e\cdot
  k}\,\Big(1 - \frac{E_e + E_{\nu} + \omega}{M}\Big)\,{\rm
  tr}\{(\hat{k}_e + m_e)\,Q_e
\gamma^0\hat{k}_{\nu}\gamma^0\,(1 - \gamma^5)\}\nonumber\\
\hspace{-0.3in}&& + \frac{1}{k_e\cdot
  k}\,\Big(\frac{\varepsilon^{0}_{\lambda'}}{\omega}\,\frac{\lambda^2
  + \lambda}{M}\,\vec{k}_p + \frac{\lambda}{M
  \omega}\,(\varepsilon^{0}_{\lambda'} \vec{k} - \omega
\vec{\varepsilon}_{\lambda'}) - \frac{\lambda}{M
  \omega}\,i\,(\vec{\varepsilon}_{\lambda'} \times
\vec{k}\,)\Big)\cdot {\rm tr}\{(\hat{k}_e + m_e)\,Q_e
\gamma^0\hat{k}_{\nu}\vec{\gamma}\,(1 - \gamma^5)\} -
\frac{1}{(k_e\cdot k)^2}\,\frac{\lambda^2 + 1}{2M}\nonumber\\
\hspace{-0.3in}&&\times\,\vec{k}_p\cdot {\rm tr}\{(\hat{k}_e +
m_e)\,Q_e \vec{\gamma}\,\hat{k}_{\nu} \gamma^0 \bar{Q}_e (1 -
\gamma^5)\} + \frac{1}{(k_e\cdot k)^2}\,\Big[\lambda^2 \Big(1 -
  \frac{E_e + E_{\nu} + \omega}{M}\Big)\,\delta^{ij} +
  \frac{\lambda\,(\kappa +
    1)}{M}\,i\,\varepsilon^{ij\ell}(\vec{k}_p)^{\ell}\Big]\nonumber\\
\hspace{-0.3in}&&\times\,{\rm tr}\{(\hat{k}_e + m_e)\,Q_e
\vec{\gamma}^{\,i}\, \hat{k}_{\nu} \vec{\gamma}^{\,j}\,
\bar{Q}_e (1 - \gamma^5)\} + \frac{1}{k_e\cdot
  k}\,\Big(\frac{\lambda^2 + 1}{M \omega}\,\varepsilon^{0}_{\lambda'}
\vec{k}_p+ \frac{\lambda^2}{M \omega}\,(\varepsilon^{0}_{\lambda'}
\vec{k} - \omega \vec{\varepsilon}_{\lambda'}) - \frac{\lambda}{M
  \omega}\,i\,(\vec{\varepsilon}_{\lambda'} \times \vec{k}\,)
\Big)\nonumber\\
\hspace{-0.3in}&&\cdot {\rm tr}\{(\hat{k}_e + m_e)\,Q_e
\vec{\gamma}\,\hat{k}_{\nu}\gamma^0 (1 - \gamma^5)\} -
2\,\frac{1}{k_e\cdot k}\,\Big[\lambda^2 \Big(1 - \frac{E_e + E_{\nu} +
    \omega}{M}\Big)\,\frac{\varepsilon^{0}_{\lambda'}}{\omega}\,
  \delta^{ij} + \frac{\lambda^2}{2 M
    \omega}\,\Big(\vec{\varepsilon}^{\,i}_{\lambda'} \vec{k}^{\,j} -
  \vec{\varepsilon}^{\,j}_{\lambda'} \vec{k}^{\,i}\Big)\nonumber\\
\hspace{-0.3in}&& + \frac{\lambda^2 + (2 \kappa + 1)\lambda}{2 M
  \omega}\,\varepsilon^{0}_{\lambda'}\,i\,\varepsilon^{ij\ell}(\vec{k}_p)^{\ell}
+ \frac{\lambda^2}{2 M \omega}\,i\,\varepsilon^{ij\ell}
\Big(\varepsilon^{0}_{\lambda'} \vec{k}^{\, \ell} - \omega
\vec{\varepsilon}^{\,\ell}_{\lambda'}\Big)\Big] {\rm tr}\{(\hat{k}_e +
m_e)\,Q_e \vec{\gamma}^{\,i}\,\hat{k}_{\nu}
\vec{\gamma}^{\,j}\,(1 - \gamma^5)\}\nonumber\\
\hspace{-0.3in}&& - 2\,\frac{1}{k_e\cdot
  k}\,\frac{\varepsilon^{0*}_{\lambda'}}{\omega}\Big(1 - \frac{E_e +
  E_{\nu} + \omega}{M}\Big)\,{\rm tr}\{(\hat{k}_e +
m_e)\,\gamma^0 \hat{k}_{\nu} \gamma^0 \bar{Q}_e \,(1 -
\gamma^5)\} + \frac{1}{k_e\cdot k}\,\Big(\frac{\lambda^2 + 1}{M
  \omega}\,\varepsilon^{0*}_{\lambda'} \vec{k}_p + \frac{\lambda^2}{M
  \omega}\,(\varepsilon^{0*}_{\lambda'} \vec{k} - \omega
\vec{\varepsilon}^{\,*}_{\lambda'})\nonumber\\
\hspace{-0.3in}&& + i\,\frac{\lambda}{M
  \omega}\,(\vec{\varepsilon}^{\,*}_{\lambda'} \times
\vec{k}\,)\Big)\cdot {\rm tr}\{(\hat{k}_e + m_e)\,\gamma^0
\,\hat{k}_{\nu} \vec{\gamma}\,\bar{Q}_e (1 - \gamma^5)\} +
4\,\frac{\varepsilon^{0*}_{\lambda'}\varepsilon^{0}_{\lambda'}}{\omega^2}\,\Big(1
- \frac{E_e + E_{\nu} + \omega}{M}\Big)\,{\rm tr}\{(\hat{k}_e
+ m_e)\,\gamma^0 \hat{k}_{\nu} \gamma^0 (1 -
\gamma^5)\}\nonumber\\
\hspace{-0.3in}&& - 2\,\Big[\frac{\lambda^2 + \lambda}{M
    \omega^2}\,\varepsilon^{0*}_{\lambda'}\varepsilon^{0}_{\lambda'}
  \vec{k}_p + \frac{\lambda^2}{M
    \omega^2}\,(\varepsilon^{0*}_{\lambda'} \vec{k} - \omega
  \vec{\varepsilon}^{\,*}_{\lambda'})\varepsilon^{0}_{\lambda'} +
  \frac{\lambda}{M
    \omega^2}\,\varepsilon^{0*}_{\lambda'}(\varepsilon^{0}_{\lambda'}
  \vec{k} - \omega \vec{\varepsilon}_{\lambda'}) + \frac{\lambda}{M
    \omega^2}\,i\,\Big((\vec{\varepsilon}^{\,*}_{\lambda'} \times
  \vec{k}\,) \varepsilon^{0}_{\lambda'} -
  \varepsilon^{0*}_{\lambda'}(\vec{\varepsilon}_{\lambda'} \times
  \vec{k}\,)\Big)\Big]\nonumber\\
\hspace{-0.3in}&&\cdot {\rm tr}\{(\hat{k}_e + m_e)\,\gamma^0
\,\hat{k}_{\nu} \vec{\gamma}\,(1 - \gamma^5)\} +
\frac{1}{k_e\cdot k}\,\Big(\frac{\lambda^2 + \lambda}{M
  \omega}\,\varepsilon^{0*}_{\lambda'}\, \vec{k}_p + \frac{\lambda}{M
  \omega}\,(\varepsilon^{0*}_{\lambda'}\, \vec{k} -
\omega\,\vec{\varepsilon}^{\,*}_{\lambda'}) + \frac{\lambda}{M
  \omega}\,i\,(\vec{\varepsilon}^{\,*}_{\lambda'} \times
\vec{k}\,)\Big)\nonumber\\
\hspace{-0.3in}&& \cdot {\rm tr}\{(\hat{k}_e +
m_e)\,\vec{\gamma}\,\hat{k}_{\nu}\gamma^0 \, \bar{Q}_e (1 -
\gamma^5)\} - 2\,\frac{1}{k_e\cdot k}\,\Big[\lambda^2 \Big(1 -
  \frac{E_e + E_{\nu} +
    \omega}{M}\Big)\,\frac{\varepsilon^{0*}_{\lambda'}}{\omega}\,
  \delta^{ij} - \frac{\lambda^2 }{2 M \omega}
  \Big(\vec{\varepsilon}^{\, i *}_{\lambda'} \vec{k}^{\,j} -
  \vec{\varepsilon}^{\,j *}_{\lambda'} \vec{k}^{\,i}\Big)\nonumber\\
\hspace{-0.3in}&& + \frac{\lambda^2 + (2 \kappa + 1)\lambda}{2 M
  \omega}\,\varepsilon^{0*}_{\lambda'}\,i\,\varepsilon^{ij\ell}
\vec{k}^{\,\ell}_p + \frac{\lambda^2}{2 M
  \omega}\,i\,\varepsilon^{ij\ell} \Big(\varepsilon^{0*}_{\lambda'}
\vec{k}^{\,\ell} - \omega \vec{\varepsilon}^{\,\ell
  *}_{\lambda'}\Big)\Big] {\rm tr}\{(\hat{k}_e + m_e)\,
\vec{\gamma}^{\,i}\,\hat{k}_{\nu} \vec{\gamma}^{\,j}\,\bar{Q}_e (1 -
\gamma^5)\}\nonumber\\
\hspace{-0.3in}&& - 2\,\Big[\frac{\lambda^2 + \lambda}{M
    \omega^2}\,\varepsilon^{0*}_{\lambda'}\varepsilon^{0}_{\lambda'}
  \vec{k}_p + \frac{\lambda^2}{M
    \omega^2}\,\varepsilon^{0*}_{\lambda'} (\varepsilon^{0}_{\lambda'}
  \vec{k} - \omega \vec{\varepsilon}_{\lambda'}) + \frac{\lambda}{M
    \omega^2}\, (\varepsilon^{0*}_{\lambda'} \vec{k} - \omega
  \vec{\varepsilon}^{\,*}_{\lambda'})\varepsilon^{0}_{\lambda'} -
  \frac{\lambda}{M \omega^2}\,i\,\Big( \varepsilon^{0*}_{\lambda'}
  (\vec{\varepsilon}_{\lambda'} \times \vec{k}\,) -
  (\vec{\varepsilon}^{\,*}_{\lambda'} \times
  \vec{k}\,)\varepsilon^{0}_{\lambda'}\Big)\Big]\nonumber\\
\hspace{-0.3in}&&\cdot {\rm tr}\{(\hat{k}_e + m_e)\,\vec{\gamma}\,
\hat{k}_{\nu} \gamma^0 (1 - \gamma^5)\} + 4\,\Big\{\lambda^2
\frac{\varepsilon^{0*}_{\lambda'}\varepsilon^{0}_{\lambda'}}{\omega^2}\,\Big(1
- \frac{E_e + E_{\nu} + \omega}{M}\Big)\, \delta^{ij} +
i\,\frac{\lambda^2 + \kappa \lambda}{M
  \omega^2}\,\varepsilon^{0*}_{\lambda'}\varepsilon^{0}_{\lambda'}\,\varepsilon^{ij\ell}(\vec{k}_p)^{\ell}\nonumber\\
\hspace{-0.3in}&& - \frac{\lambda^2}{2 M
  \omega^2}\,\Big[\Big(\vec{\varepsilon}^{\,i *}_{\lambda'}
  \vec{k}^{\,j} - \vec{\varepsilon}^{\,j *}_{\lambda'}
  \vec{k}^{\,i}\Big)\varepsilon^{0}_{\lambda'} -
  \varepsilon^{0*}_{\lambda'}\Big(\vec{\varepsilon}^{\,i}_{\lambda'}
  \vec{k}^{\,j} - \vec{\varepsilon}^{\,j}_{\lambda'}
  \vec{k}^{\,i}\Big)\Big]\nonumber\\
\hspace{-0.3in}&& + \frac{\lambda^2}{2 M
  \omega^2}\,i\,\varepsilon^{ij\ell}\Big[\Big(\varepsilon^{0*}_{\lambda'}
  \vec{k}^{\,\ell} - \omega \vec{\varepsilon}^{\,\ell
    *}_{\lambda'}\Big) \varepsilon^{0}_{\lambda'} +
  \varepsilon^{0*}_{\lambda'}\Big(\varepsilon^{0}_{\lambda'}
  \vec{k}^{\,\ell} - \omega
  \vec{\varepsilon}^{\,\ell}_{\lambda'}\Big)\Big]\Big\}\,{\rm
  tr}\{(\hat{k}_e + m_e) \vec{\gamma}^{\,i} \hat{k}_{\nu}
\vec{\gamma}^{\,j} (1 - \gamma^5)\}.
\end{eqnarray}
One may show that the expression in Eq.(\ref{eq:A.21}) is gauge
invariant. For this aim one has to replace $\varepsilon^*_{\lambda'}
\to k$ and $\varepsilon_{\lambda'} \to k$ and collect the coefficients
in font of the traces ${\rm tr}\{(\hat{k}_e + m_e) \gamma^{\alpha}
\hat{k}_{\nu} \gamma^{\beta}(1 - \gamma^5)\}$.  Now we may proceed to
the calculation of the traces over electron and antineutrino degrees
of freedom. Using the following relations \cite{Ivanov2013} (see
Appendix B of Ref.\cite{Ivanov2013})
\begin{eqnarray}\label{eq:A.22}
\hspace{-0.3in}\gamma^{\alpha}\gamma^{\nu}\gamma^{\mu} =
\gamma^{\alpha}\eta^{\nu\mu} - \gamma^{\nu}\eta^{\mu\alpha} +
\gamma^{\mu}\eta^{\alpha\nu} +
i\,\varepsilon^{\alpha\nu\mu\beta}\,\gamma_{\beta}\gamma^5,
\end{eqnarray}
where $\eta^{\alpha\beta}$ is the Minkowski metric tensor and
$\varepsilon^{\alpha\nu\mu\beta}$ is the Levi--Civita tensor defined
by $\varepsilon^{0123} = 1$ and $\varepsilon_{\alpha\nu\mu\beta}= -
\varepsilon^{\alpha\nu\mu\beta}$ \cite{IZ1980} and \cite{Ivanov2013}
(see Appendix B of Ref.\cite{Ivanov2013}), we obtain
\begin{eqnarray*}
\hspace{-0.3in}&&\frac{1}{4}\,{\rm tr}\{\hat{k}_e Q_e \gamma^{\mu}
\bar{Q}_e(1 - \gamma^5)\} = 4(\varepsilon^*_{\lambda'}\cdot
k_e)(\varepsilon_{\lambda'}\cdot k_e)\,(k_e + k)^{\mu} -
2\,(\varepsilon^*_{\lambda'}\cdot
\varepsilon_{\lambda'})\,(k_e\cdot k)\,k^{\mu}\nonumber\\
\end{eqnarray*}
\begin{eqnarray}\label{eq:A.23}
\hspace{-0.3in}&& - 2\,\Big((\varepsilon^*_{\lambda'}\cdot
k_e)\,\varepsilon^{\mu}_{\lambda'} + (\varepsilon_{\lambda'} \cdot
k_e)\,\varepsilon^{\mu *}_{\lambda'}\Big)\,(k_e\cdot k) +
2\,i\,\varepsilon^{\mu\alpha\beta\nu}\Big((\varepsilon^*_{\lambda'}\cdot
k_e)\,\varepsilon_{\lambda' \alpha} - (\varepsilon_{\lambda'} \cdot
k_e)\,\varepsilon^*_{\lambda'\alpha}\Big)\,k_{\beta} k_{e
  \nu}\nonumber\\
\hspace{-0.3in}&& + 2\,i\,k^{\mu}\, \varepsilon^{
  \alpha\beta\rho\nu}\,\varepsilon^*_{\lambda'\alpha}\,
\varepsilon_{\lambda'\beta}\,k_{\rho}\,k_{e\nu}
\end{eqnarray}
for a photon on--mass shell $k^2 = 0$. Then, we get
\begin{eqnarray}\label{eq:A.24}
\hspace{-0.3in}&&\frac{1}{4}\,{\rm tr}\{\hat{k}_e Q_e \gamma^{\mu} (1
- \gamma^5)\} = (\varepsilon^*_{\lambda'}\cdot k_e) (2 k_e + k)^{\mu}
- (k_e \cdot k) \varepsilon^{\mu *}_{\lambda'} -
i\,\varepsilon^{\mu\alpha\beta\nu} \varepsilon^*_{\lambda' \alpha}
k_{\beta} k_{e\nu},\nonumber\\
\hspace{-0.3in}&&\frac{1}{4}\,{\rm tr}\{\hat{k}_e \gamma^{\mu}
\bar{Q}_e (1 - \gamma^5)\} = (\varepsilon_{\lambda'}\cdot k_e) (2 k_e
+ k)^{\mu} - (k_e \cdot k) \varepsilon^{\mu}_{\lambda'} +
i\,\varepsilon^{\mu\alpha\beta\nu} \varepsilon_{\lambda' \alpha}
k_{\beta} k_{e\nu}.
\end{eqnarray}
Using Eq.(\ref{eq:A.23}) and Eq.(\ref{eq:A.24}) for the traces over
the electron and antineutrino degrees of freedom, containing both
$Q_e$ and $\bar{Q}_e$ matrices, we obtain the following expressions
\begin{eqnarray*}
\hspace{-0.3in}&&\frac{1}{4}\,{\rm tr}\{(\hat{k}_e + m_e)\,Q_e
\gamma^0\hat{k}_{\nu}\gamma^0 \bar{Q}_e (1 - \gamma^5)\} =
E_{\nu} \Big[4(\varepsilon^*_{\lambda'}\cdot
  k_e)(\varepsilon_{\lambda'}\cdot k_e)\,(E_e + \omega) -
  2\,\omega\,(\varepsilon^*_{\lambda'}\cdot
  \varepsilon_{\lambda'})\,(k_e\cdot k)\nonumber\\
\hspace{-0.3in}&& - 2\,\Big((\varepsilon^*_{\lambda'}\cdot
k_e)\,\varepsilon^0_{\lambda'} + (\varepsilon_{\lambda'} \cdot
k_e)\,\varepsilon^{0}_{\lambda'}\Big)\,(k_e\cdot k)  +
2\,i\,\varepsilon^{0\alpha\beta\nu}\Big((\varepsilon^*_{\lambda'}\cdot
k_e)\,\varepsilon_{\lambda' \alpha} - (\varepsilon_{\lambda'} \cdot
k_e)\,\varepsilon^*_{\lambda'\alpha}\Big)\,k_{\beta} k_{e
  \nu}\nonumber\\
\hspace{-0.3in}&& + 2\,i\,\omega\, \varepsilon^{
  \alpha\beta\rho\nu}\,\varepsilon^*_{\lambda'\alpha}\,
\varepsilon_{\lambda'\beta}\,k_{\rho}\,k_{e\nu}\Big] +
\vec{k}^{\,i}_{\nu} \Big[4(\varepsilon^*_{\lambda'}\cdot k_e)(\varepsilon_{\lambda'}\cdot
k_e)\,(\vec{k}_e + \vec{k}\,)^i -
2\,(\varepsilon^*_{\lambda'}\cdot \varepsilon_{\lambda'})\,(k_e\cdot
k)\,\vec{k}^{\,i}\nonumber\\
\hspace{-0.3in}&& - 2\,\Big((\varepsilon^*_{\lambda'}\cdot
k_e)\,(\vec{\varepsilon}^{\,i}_{\lambda'} + (\varepsilon_{\lambda'}
\cdot k_e)\, \vec{\varepsilon}^{\,i *}_{\lambda'}\Big)\,(k_e\cdot k) +
2\,i\,\varepsilon^{i\alpha\beta\nu}\Big((\varepsilon^*_{\lambda'}\cdot
k_e)\,\varepsilon_{\lambda' \alpha} - (\varepsilon_{\lambda'} \cdot
k_e)\,\varepsilon^*_{\lambda'\alpha}\Big)\,k_{\beta} k_{e
  \nu}\nonumber\\
\hspace{-0.3in}&& + 2\,i\,\vec{k}^{\,i}\, \varepsilon^{
  \alpha\beta\rho\nu}\,\varepsilon^*_{\lambda'\alpha}\,
\varepsilon_{\lambda'\beta}\,k_{\rho}\,k_{e\nu}\Big],\nonumber\\
\hspace{-0.3in}&&\frac{1}{4}\,{\rm tr}\{(\hat{k}_e + m_e)\,Q_e
\gamma^0\hat{k}_{\nu}\vec{\gamma}^{\,i}\, \bar{Q}_e (1 -
\gamma^5)\} = E_{\nu}\Big[4(\varepsilon^*_{\lambda'}\cdot
k_e)(\varepsilon_{\lambda'}\cdot k_e)\,(\vec{k}_e + \vec{k}\,)^i
 - 2\,(\varepsilon^*_{\lambda'}\cdot
\varepsilon_{\lambda'})\,(k_e\cdot k)\,\vec{k}^{\,i}
\nonumber\\
\hspace{-0.3in}&& - 2\,\Big((\varepsilon^*_{\lambda'}\cdot
k_e)\,\vec{\varepsilon}^{\,i}_{\lambda'} + (\varepsilon_{\lambda'} \cdot
k_e)\,\vec{\varepsilon}^{\,i *}_{\lambda'}\Big)\,(k_e\cdot k)  +
2\,i\,\varepsilon^{i \alpha\beta\nu}\Big((\varepsilon^*_{\lambda'}\cdot
k_e)\,\varepsilon_{\lambda' \alpha} - (\varepsilon_{\lambda'} \cdot
k_e)\,\varepsilon^*_{\lambda'\alpha}\Big)\,k_{\beta} k_{e
  \nu}\nonumber\\
\hspace{-0.3in}&& + 2\,i\,\vec{k}^{\,i}\, \varepsilon^{
  \alpha\beta\rho\nu}\,\varepsilon^*_{\lambda'\alpha}\,
\varepsilon_{\lambda'\beta}\,k_{\rho}\,k_{e\nu}\Big] +
\vec{k}^{\,i}_{\nu}\Big[4(\varepsilon^*_{\lambda'}\cdot
  k_e)(\varepsilon_{\lambda'}\cdot k_e)\,(E_e + \omega) -
  2\,\omega\,(\varepsilon^*_{\lambda'}\cdot
  \varepsilon_{\lambda'})\,(k_e\cdot k)\nonumber\\
\hspace{-0.3in}&& - 2\,\Big((\varepsilon^*_{\lambda'}\cdot
k_e)\,\varepsilon^0_{\lambda'} + (\varepsilon_{\lambda'} \cdot
k_e)\,\varepsilon^{0}_{\lambda'}\Big)\,(k_e\cdot k)  +
2\,i\,\varepsilon^{0\alpha\beta\nu}\Big((\varepsilon^*_{\lambda'}\cdot
k_e)\,\varepsilon_{\lambda' \alpha} - (\varepsilon_{\lambda'} \cdot
k_e)\,\varepsilon^*_{\lambda'\alpha}\Big)\,k_{\beta} k_{e
  \nu}\nonumber\\
\hspace{-0.3in}&& + 2\,i\,\omega\, \varepsilon^{
  \alpha\beta\rho\nu}\,\varepsilon^*_{\lambda'\alpha}\,
\varepsilon_{\lambda'\beta}\,k_{\rho}\,k_{e\nu}\Big] +
i\,\varepsilon^{ij\ell} \vec{k}^{\,j}_{\nu}
\Big[4(\varepsilon^*_{\lambda'}\cdot k_e)(\varepsilon_{\lambda'}\cdot
  k_e)\,(\vec{k}_e + \vec{k}\,)^{\ell} -
  2\,(\varepsilon^*_{\lambda'}\cdot \varepsilon_{\lambda'})\,(k_e\cdot
  k)\,(\vec{k}\,)^{\ell}\nonumber\\
\hspace{-0.3in}&& - 2\,\Big((\varepsilon^*_{\lambda'}\cdot
k_e)\,(\vec{\varepsilon}^{\,\ell}_{\lambda'} +
(\varepsilon_{\lambda'} \cdot k_e)\,(
\vec{\varepsilon}^{\,\ell  *}_{\lambda'}\Big)\,(k_e\cdot k) +
2\,i\,\varepsilon^{\ell \alpha\beta\nu}\Big((\varepsilon^*_{\lambda'}\cdot
k_e)\,\varepsilon_{\lambda' \alpha} - (\varepsilon_{\lambda'} \cdot
k_e)\,\varepsilon^*_{\lambda'\alpha}\Big)\,k_{\beta} k_{e
  \nu}\nonumber\\
\hspace{-0.3in}&& + 2\,i\,\vec{k}^{\,\ell}\, \varepsilon^{
  \alpha\beta\rho\nu}\,\varepsilon^*_{\lambda'\alpha}\,
\varepsilon_{\lambda'\beta}\,k_{\rho}\,k_{e\nu}\Big],\nonumber\\
\hspace{-0.3in}&&\frac{1}{4}\,{\rm tr}\{(\hat{k}_e + m_e)\,Q_e
\vec{\gamma}^{\,i}\,\hat{k}_{\nu} \gamma^0\,\bar{Q}_e (1 -
\gamma^5)\} = E_{\nu}\Big[4(\varepsilon^*_{\lambda'}\cdot k_e)(\varepsilon_{\lambda'}\cdot
k_e)\,(\vec{k}_e + \vec{k}\,)^i -
2\,(\varepsilon^*_{\lambda'}\cdot \varepsilon_{\lambda'})\,(k_e\cdot
k)\,\vec{k}^{\,i}\nonumber\\
\hspace{-0.3in}&& - 2\,\Big((\varepsilon^*_{\lambda'}\cdot
k_e)\,(\vec{\varepsilon}^{\,i}_{\lambda'} +
(\varepsilon_{\lambda'} \cdot k_e)\,(
\vec{\varepsilon}^{\, i *}_{\lambda'}\Big)\,(k_e\cdot k) +
2\,i\,\varepsilon^{i\alpha\beta\nu}\Big((\varepsilon^*_{\lambda'}\cdot
k_e)\,\varepsilon_{\lambda' \alpha} - (\varepsilon_{\lambda'} \cdot
k_e)\,\varepsilon^*_{\lambda'\alpha}\Big)\,k_{\beta} k_{e
  \nu}\nonumber\\
\hspace{-0.3in}&& + 2\,i\,\vec{k}^{\,i}\, \varepsilon^{
  \alpha\beta\rho\nu}\,\varepsilon^*_{\lambda'\alpha}\,
\varepsilon_{\lambda'\beta}\,k_{\rho}\,k_{e\nu}\Big] +
\vec{k}^{\,i}_{\nu} \Big[4(\varepsilon^*_{\lambda'}\cdot
  k_e)(\varepsilon_{\lambda'}\cdot k_e)\,(E_e + \omega) -
  2\,\omega\,(\varepsilon^*_{\lambda'}\cdot
  \varepsilon_{\lambda'})\,(k_e\cdot k)\nonumber\\
\hspace{-0.3in}&& - 2\,\Big((\varepsilon^*_{\lambda'}\cdot
k_e)\,\varepsilon^0_{\lambda'} + (\varepsilon_{\lambda'} \cdot
k_e)\,\varepsilon^{0}_{\lambda'}\Big)\,(k_e\cdot k)  +
2\,i\,\varepsilon^{0\alpha\beta\nu}\Big((\varepsilon^*_{\lambda'}\cdot
k_e)\,\varepsilon_{\lambda' \alpha} - (\varepsilon_{\lambda'} \cdot
k_e)\,\varepsilon^*_{\lambda'\alpha}\Big)\,k_{\beta} k_{e
  \nu}\nonumber\\
\hspace{-0.3in}&& + 2\,i\,\omega\, \varepsilon^{
  \alpha\beta\rho\nu}\,\varepsilon^*_{\lambda'\alpha}\,
\varepsilon_{\lambda'\beta}\,k_{\rho}\,k_{e\nu}\Big] -
i\,\varepsilon^{ij\ell} \vec{k}^{\,j}_{\nu}
\Big[4(\varepsilon^*_{\lambda'}\cdot k_e)(\varepsilon_{\lambda'}\cdot
  k_e)\,(\vec{k}_e + \vec{k}\,)^{\ell} -
  2\,(\varepsilon^*_{\lambda'}\cdot \varepsilon_{\lambda'})\,(k_e\cdot
  k)\,\vec{k}^{\,\ell}\nonumber\\
\hspace{-0.3in}&& - 2\,\Big((\varepsilon^*_{\lambda'}\cdot
k_e)\,\vec{\varepsilon}^{\,\ell}_{\lambda'} +
(\varepsilon_{\lambda'} \cdot k_e)\,
\vec{\varepsilon}^{\,\ell *}_{\lambda'}\Big)\,(k_e\cdot k) +
2\,i\,\varepsilon^{\ell \alpha\beta\nu}\Big((\varepsilon^*_{\lambda'}\cdot
k_e)\,\varepsilon_{\lambda' \alpha} - (\varepsilon_{\lambda'} \cdot
k_e)\,\varepsilon^*_{\lambda'\alpha}\Big)\,k_{\beta} k_{e
  \nu}\nonumber\\
\hspace{-0.3in}&& + 2\,i\,\vec{k}^{\,\ell}\, \varepsilon^{
  \alpha\beta\rho\nu}\,\varepsilon^*_{\lambda'\alpha}\,
\varepsilon_{\lambda'\beta}\,k_{\rho}\,k_{e\nu}\Big],\nonumber\\
\hspace{-0.3in}&&\frac{1}{4}\,{\rm tr}\{(\hat{k}_e + m_e)\,Q_e
\vec{\gamma}^{\,i}\,\hat{k}_{\nu}
\vec{\gamma}^{\,j}\,\bar{Q}_e (1 - \gamma^5)\} =\nonumber\\
\hspace{-0.3in}&&= E_{\nu}\frac{1}{4}\,{\rm tr}\{\hat{k}_e \,Q_e
\vec{\gamma}^{\,i}\,\gamma^0 \vec{\gamma}^{\,j}\,\bar{Q}_e (1 -
\gamma^5)\} - \vec{k}^{\,\ell}_{\nu} \frac{1}{4}\,{\rm
  tr}\{\hat{k}_e\,Q_e \vec{\gamma}^{\,i}\,\vec{\gamma}^{\,\ell}
\vec{\gamma}^{\,j}\,\bar{Q}_e (1 - \gamma^5)\}=\nonumber\\
\hspace{-0.3in}&&= \delta^{ij}\,E_{\nu}
\Big[4(\varepsilon^*_{\lambda'}\cdot k_e)(\varepsilon_{\lambda'}\cdot
  k_e)\,(E_e + \omega) - 2\,\omega\,(\varepsilon^*_{\lambda'}\cdot
  \varepsilon_{\lambda'})\,(k_e\cdot k) -
  2\,\Big((\varepsilon^*_{\lambda'}\cdot
  k_e)\,\varepsilon^0_{\lambda'} + (\varepsilon_{\lambda'} \cdot
  k_e)\,\varepsilon^{0}_{\lambda'}\Big)\,(k_e\cdot k)\nonumber\\
\hspace{-0.3in}&& +
2\,i\,\varepsilon^{0\alpha\beta\nu}\Big((\varepsilon^*_{\lambda'}\cdot
k_e)\,\varepsilon_{\lambda' \alpha} - (\varepsilon_{\lambda'} \cdot
k_e)\,\varepsilon^*_{\lambda'\alpha}\Big)\,k_{\beta} k_{e
  \nu} + 2\,i\,\omega\, \varepsilon^{
  \alpha\beta\rho\nu}\,\varepsilon^*_{\lambda'\alpha}\,
\varepsilon_{\lambda'\beta}\,k_{\rho}\,k_{e\nu}\Big] -
i\,\varepsilon^{ij\ell} E_{\nu} \nonumber\\
\end{eqnarray*}
\begin{eqnarray}\label{eq:A.25}
\hspace{-0.3in}&&\times \Big[4(\varepsilon^*_{\lambda'}\cdot
  k_e)(\varepsilon_{\lambda'}\cdot k_e)\,(\vec{k}_e +
  \vec{k}\,)^{\ell} - 2\,(\varepsilon^*_{\lambda'}\cdot
  \varepsilon_{\lambda'})\,(k_e\cdot k)\,\vec{k}^{\,\ell} -
  2\,\Big((\varepsilon^*_{\lambda'}\cdot
  k_e)\,\vec{\varepsilon}^{\,\ell}_{\lambda'} +
  (\varepsilon_{\lambda'} \cdot k_e)\,\vec{\varepsilon}^{\,
    \ell *}_{\lambda'}\Big)\,(k_e\cdot k)\nonumber\\
\hspace{-0.3in}&& + 2\,i\,\varepsilon^{\ell
  \alpha\beta\nu}\Big((\varepsilon^*_{\lambda'}\cdot
k_e)\,\varepsilon_{\lambda' \alpha} - (\varepsilon_{\lambda'} \cdot
k_e)\,\varepsilon^*_{\lambda'\alpha}\Big)\,k_{\beta} k_{e \nu} +
2\,i\,(\vec{k}\,)^{\ell}\, \varepsilon^{
  \alpha\beta\rho\nu}\,\varepsilon^*_{\lambda'\alpha}\,
\varepsilon_{\lambda'\beta}\,k_{\rho}\,k_{e\nu}\Big] - \delta^{ij}\,
\vec{k}^{\,\ell}_{\nu} \nonumber\\
\hspace{-0.3in}&&\times \Big[4(\varepsilon^*_{\lambda'}\cdot
  k_e)(\varepsilon_{\lambda'}\cdot k_e)\,(\vec{k}_e +
  \vec{k}\,)^{\ell} - 2\,(\varepsilon^*_{\lambda'}\cdot
  \varepsilon_{\lambda'})\,(k_e\cdot k)\,\vec{k}^{\,\ell} -
  2\,\Big((\varepsilon^*_{\lambda'}\cdot
  k_e)\,\vec{\varepsilon}^{\,\ell}_{\lambda'} +
  (\varepsilon_{\lambda'} \cdot k_e)\,\vec{\varepsilon}^{\,
    \ell *}_{\lambda'}\Big)\,(k_e\cdot k)\nonumber\\
\hspace{-0.3in}&& + 2\,i\,\varepsilon^{\ell
  \alpha\beta\nu}\Big((\varepsilon^*_{\lambda'}\cdot
k_e)\,\varepsilon_{\lambda' \alpha} - (\varepsilon_{\lambda'} \cdot
k_e)\,\varepsilon^*_{\lambda'\alpha}\Big)\,k_{\beta} k_{e \nu} +
2\,i\,\vec{k}^{\,\ell}\, \varepsilon^{
  \alpha\beta\rho\nu}\,\varepsilon^*_{\lambda'\alpha}\,
\varepsilon_{\lambda'\beta}\,k_{\rho}\,k_{e\nu}\Big] +
\vec{k}^{\,i}_{\nu} \nonumber\\
\hspace{-0.3in}&& \times \Big[4(\varepsilon^*_{\lambda'}\cdot
  k_e)(\varepsilon_{\lambda'}\cdot k_e)\,(\vec{k}_e +
  \vec{k}\,)^j - 2\,(\varepsilon^*_{\lambda'}\cdot
  \varepsilon_{\lambda'})\,(k_e\cdot k)\,\vec{k}^{\,j} -
  2\,\Big((\varepsilon^*_{\lambda'}\cdot
  k_e)\,\vec{\varepsilon}^{\,j}_{\lambda'} +
  (\varepsilon_{\lambda'} \cdot k_e)\,\vec{\varepsilon}^{\,
    j *}_{\lambda'}\Big)\,(k_e\cdot k)\nonumber\\
\hspace{-0.3in}&& + 2\,i\,\varepsilon^{j
  \alpha\beta\nu}\Big((\varepsilon^*_{\lambda'}\cdot
k_e)\,\varepsilon_{\lambda' \alpha} - (\varepsilon_{\lambda'} \cdot
k_e)\,\varepsilon^*_{\lambda'\alpha}\Big)\,k_{\beta} k_{e \nu} +
2\,i\,\vec{k}^{\,j}\, \varepsilon^{
  \alpha\beta\rho\nu}\,\varepsilon^*_{\lambda'\alpha}\,
\varepsilon_{\lambda'\beta}\,k_{\rho}\,k_{e\nu}\Big] + \vec{k}^{\,j}_{\nu}\nonumber\\
\hspace{-0.3in}&& \times \Big[4(\varepsilon^*_{\lambda'}\cdot
  k_e)(\varepsilon_{\lambda'}\cdot k_e)\,(\vec{k}_e +
  \vec{k}\,)^i - 2\,(\varepsilon^*_{\lambda'}\cdot
  \varepsilon_{\lambda'})\,(k_e\cdot k)\,\vec{k}^{\,i} -
  2\,\Big((\varepsilon^*_{\lambda'}\cdot
  k_e)\,\vec{\varepsilon}^{\,i}_{\lambda'} +
  (\varepsilon_{\lambda'} \cdot k_e)\,\vec{\varepsilon}^{\,
   i *}_{\lambda'}\Big)\,(k_e\cdot k)\nonumber\\
\hspace{-0.3in}&& + 2\,i\,\varepsilon^{i
  \alpha\beta\nu}\Big((\varepsilon^*_{\lambda'}\cdot
k_e)\,\varepsilon_{\lambda' \alpha} - (\varepsilon_{\lambda'} \cdot
k_e)\,\varepsilon^*_{\lambda'\alpha}\Big)\,k_{\beta} k_{e \nu} +
2\,i\,\vec{k}^{\,i}\, \varepsilon^{
  \alpha\beta\rho\nu}\,\varepsilon^*_{\lambda'\alpha}\,
\varepsilon_{\lambda'\beta}\,k_{\rho}\,k_{e\nu}\Big] -
i\,\varepsilon^{ij\ell} \vec{k}^{\,\ell}_{\nu}\nonumber\\
\hspace{-0.3in}&&\times \Big[4(\varepsilon^*_{\lambda'}\cdot
  k_e)(\varepsilon_{\lambda'}\cdot k_e)\,(E_e + \omega) -
  2\,\omega\,(\varepsilon^*_{\lambda'}\cdot
  \varepsilon_{\lambda'})\,(k_e\cdot k) - 2\,\Big((\varepsilon^*_{\lambda'}\cdot
k_e)\,\varepsilon^0_{\lambda'} + (\varepsilon_{\lambda'} \cdot
k_e)\,\varepsilon^{0}_{\lambda'}\Big)\,(k_e\cdot k)\nonumber\\
\hspace{-0.3in}&& +
2\,i\,\varepsilon^{0\alpha\beta\nu}\Big((\varepsilon^*_{\lambda'}\cdot
k_e)\,\varepsilon_{\lambda' \alpha} - (\varepsilon_{\lambda'} \cdot
k_e)\,\varepsilon^*_{\lambda'\alpha}\Big)\,k_{\beta} k_{e \nu} +
2\,i\,\omega\, \varepsilon^{
  \alpha\beta\rho\nu}\,\varepsilon^*_{\lambda'\alpha}\,
\varepsilon_{\lambda'\beta}\,k_{\rho}\,k_{e\nu}\Big]
\end{eqnarray}
For the traces over the electron and antineutrino degrees of freedom,
containing either $Q_e$ or $\bar{Q}_e$, we get
\begin{eqnarray}\label{eq:A.26}
\hspace{-0.3in}&&\frac{1}{4}\,{\rm tr}\{(\hat{k}_e + m_e)\,Q_e
\gamma^0\hat{k}_{\nu}\gamma^0 (1 - \gamma^5)\} =
E_{\nu}\Big[ (\varepsilon^*_{\lambda'}\cdot k_e) (2 E_e +
  \omega) - (k_e \cdot k) \varepsilon^{0*}_{\lambda'} -
  i\,\varepsilon^{0 \alpha\beta\nu} \varepsilon^*_{\lambda' \alpha}
  k_{\beta} k_{e\nu}\Big]\nonumber\\
\hspace{-0.3in}&&+ \vec{k}^{\,i}_{\nu} \Big[
  (\varepsilon^*_{\lambda'}\cdot k_e) (2 \vec{k}_e + \vec{k})^i - (k_e
  \cdot k) \vec{\varepsilon}^{\,i *}_{\lambda'} - i\,\varepsilon^{i
    \alpha\beta\nu} \varepsilon^*_{\lambda' \alpha} k_{\beta}
  k_{e\nu}\Big],\nonumber\\
\hspace{-0.3in}&&\frac{1}{4}\,{\rm tr}\{(\hat{k}_e + m_e)\,Q_e
\gamma^0\hat{k}_{\nu}\vec{\gamma}^{\,i} (1 - \gamma^5)\} =
E_{\nu} \Big[ (\varepsilon^*_{\lambda'}\cdot k_e) (2 \vec{k}_e
  + \vec{k})^i - (k_e \cdot k) \vec{\varepsilon}^{\,i *}_{\lambda'} -
  i\,\varepsilon^{i \alpha\beta\nu} \varepsilon^*_{\lambda' \alpha}
  k_{\beta} k_{e\nu}\Big]\nonumber\\
\hspace{-0.3in}&&+ \vec{k}^{\,i}_{\nu}\Big[
  (\varepsilon^*_{\lambda'}\cdot k_e) (2 E_e + \omega) - (k_e \cdot k)
  \varepsilon^{0*}_{\lambda'} - i\,\varepsilon^{0 \alpha\beta\nu}
  \varepsilon^*_{\lambda' \alpha} k_{\beta} k_{e\nu}\Big]\nonumber\\
\hspace{-0.3in}&&+ i\,\varepsilon^{ij\ell} \vec{k}^{\,j}_{\nu}\Big[
  (\varepsilon^*_{\lambda'}\cdot k_e) (2 \vec{k}_e + \vec{k})^{\ell} -
  (k_e \cdot k) \vec{\varepsilon}^{\,\ell *}_{\lambda'} -
  i\,\varepsilon^{\ell \alpha\beta\nu} \varepsilon^*_{\lambda' \alpha}
  k_{\beta} k_{e\nu}\Big],\nonumber\\
\hspace{-0.3in}&&\frac{1}{4}\,{\rm tr}\{(\hat{k}_e + m_e)\,Q_e
\vec{\gamma}^{\,i} \hat{k}_{\nu}\gamma^0 (1 - \gamma^5)\} =
E_{\nu} \Big[ (\varepsilon^*_{\lambda'}\cdot k_e) (2 \vec{k}_e
  + \vec{k})^i - (k_e \cdot k) \vec{\varepsilon}^{\, i *}_{\lambda'} -
  i\,\varepsilon^{i \alpha\beta\nu} \varepsilon^*_{\lambda' \alpha}
  k_{\beta} k_{e\nu}\Big]\nonumber\\
\hspace{-0.3in}&&+ \vec{k}^{\,i}_{\nu} \Big[
  (\varepsilon^*_{\lambda'}\cdot k_e) (2 E_e + \omega) - (k_e \cdot k)
  \varepsilon^{0*}_{\lambda'} - i\,\varepsilon^{0 \alpha\beta\nu}
  \varepsilon^*_{\lambda' \alpha} k_{\beta} k_{e\nu}\Big]\nonumber\\
\hspace{-0.3in}&&-
i\,\varepsilon^{ij\ell} \vec{k}^{\,j}_{\nu} \Big[
  (\varepsilon^*_{\lambda'}\cdot k_e) (2 \vec{k}_e + \vec{k})^{\ell} -
  (k_e \cdot k)\vec{\varepsilon}^{\,\ell *}_{\lambda'} - i\,\varepsilon^{\ell
    \alpha\beta\nu} \varepsilon^*_{\lambda' \alpha} k_{\beta}
  k_{e\nu}\Big],\nonumber\\
\hspace{-0.3in}&&\frac{1}{4}\,{\rm tr}\{(\hat{k}_e + m_e)\,Q_e
\vec{\gamma}^{\,i} \hat{k}_{\nu}\vec{\gamma}^{\,j} (1 - \gamma^5)\} = \delta^{ij}\,E_{\nu}\Big[ (\varepsilon^*_{\lambda'}\cdot k_e) (2 E_e +
  \omega) - (k_e \cdot k) \varepsilon^{0*}_{\lambda'} -
  i\,\varepsilon^{0 \alpha\beta\nu} \varepsilon^*_{\lambda' \alpha}
  k_{\beta} k_{e\nu}\Big]\nonumber\\
\hspace{-0.3in}&& - i\,\varepsilon^{ij\ell}\,E_{\nu}\Big[
  (\varepsilon^*_{\lambda'}\cdot k_e) (2 \vec{k}_e + \vec{k})^{\ell} -
  (k_e \cdot k) \vec{\varepsilon}^{\, \ell *}_{\lambda'} - i\,\varepsilon^{\ell
    \alpha\beta\nu} \varepsilon^*_{\lambda' \alpha} k_{\beta}
  k_{e\nu}\Big]\nonumber\\
\hspace{-0.3in}&& - \delta^{ij}\,\vec{k}^{\,\ell}_{\nu} \Big[
    (\varepsilon^*_{\lambda'}\cdot k_e) (2 \vec{k}_e + \vec{k})^{\ell}
    - (k_e \cdot k) \vec{\varepsilon}^{\, \ell *}_{\lambda'} -
    i\,\varepsilon^{\ell \alpha\beta\nu} \varepsilon^*_{\lambda'
      \alpha} k_{\beta} k_{e\nu}\Big]\nonumber\\
\hspace{-0.3in}&& + \vec{k}^{\,i}_{\nu} \Big[
  (\varepsilon^*_{\lambda'}\cdot k_e) (2 \vec{k}_e + \vec{k})^j - (k_e
  \cdot k) \vec{\varepsilon}^{\,j *}_{\lambda'} - i\,\varepsilon^{j
    \alpha\beta\nu} \varepsilon^*_{\lambda' \alpha} k_{\beta}
  k_{e\nu}\Big]\nonumber\\
\hspace{-0.3in}&& + \vec{k}^{\,j}_{\nu} \Big[
  (\varepsilon^*_{\lambda'}\cdot k_e) (2 \vec{k}_e + \vec{k})^i - (k_e
  \cdot k) \vec{\varepsilon}^{\,i *}_{\lambda'} - i\,\varepsilon^{i
    \alpha\beta\nu} \varepsilon^*_{\lambda' \alpha} k_{\beta}
  k_{e\nu}\Big] \nonumber\\
\hspace{-0.3in}&& - i\,\varepsilon^{ij\ell}\,\vec{k}^{\,\ell}_{\nu}
\Big[ (\varepsilon^*_{\lambda'}\cdot k_e) (2 E_e + \omega) - (k_e
  \cdot k) \varepsilon^{0*}_{\lambda'} - i\,\varepsilon^{0
    \alpha\beta\nu} \varepsilon^*_{\lambda' \alpha} k_{\beta}
  k_{e\nu}\Big]
\end{eqnarray}
and 
\begin{eqnarray*}
\hspace{-0.3in}&&\frac{1}{4}\,{\rm tr}\{(\hat{k}_e + m_e)\,
\gamma^0\hat{k}_{\nu}\gamma^0 \bar{Q}_e(1 - \gamma^5)\} =
E_{\nu}\Big[ (\varepsilon_{\lambda'}\cdot k_e) (2 E_e +
  \omega) - (k_e \cdot k) \varepsilon^{0}_{\lambda'} +
  i\,\varepsilon^{0 \alpha\beta\nu} \varepsilon_{\lambda' \alpha}
  k_{\beta} k_{e\nu}\Big]\nonumber\\
\hspace{-0.3in}&&+ \vec{k}^{\,i}_{\nu} \Big[
  (\varepsilon_{\lambda'}\cdot k_e) (2 \vec{k}_e + \vec{k})^i - (k_e
  \cdot k) \vec{\varepsilon}^{\,i}_{\lambda'} + i\,\varepsilon^{i
    \alpha\beta\nu} \varepsilon_{\lambda' \alpha} k_{\beta}
  k_{e\nu}\Big],\nonumber\\
\hspace{-0.3in}&&\frac{1}{4}\,{\rm tr}\{(\hat{k}_e + m_e)\,
\gamma^0\hat{k}_{\nu}\vec{\gamma}^{\,i} \bar{Q}_e(1 - \gamma^5)\} =
E_{\nu} \Big[ (\varepsilon_{\lambda'}\cdot k_e) (2 \vec{k}_e
  + \vec{k})^i - (k_e \cdot k) \vec{\varepsilon}^{\,i}_{\lambda'} +
  i\,\varepsilon^{i \alpha\beta\nu} \varepsilon_{\lambda' \alpha}
  k_{\beta} k_{e\nu}\Big]\nonumber\\
\hspace{-0.3in}&&+ \vec{k}^{\,i}_{\nu} \Big[
  (\varepsilon_{\lambda'}\cdot k_e) (2 E_e + \omega) - (k_e \cdot k)
  \varepsilon^{0}_{\lambda'} + i\,\varepsilon^{0 \alpha\beta\nu}
  \varepsilon_{\lambda' \alpha} k_{\beta} k_{e\nu}\Big]\nonumber\\
\end{eqnarray*}
\begin{eqnarray}\label{eq:A.27}
\hspace{-0.3in}&& -
i\,\varepsilon^{ij\ell} \vec{k}^{\,j}_{\nu} \Big[
  (\varepsilon_{\lambda'}\cdot k_e) (2 \vec{k}_e + \vec{k})^{\ell} -
  (k_e \cdot k) \vec{\varepsilon}^{\,\ell }_{\lambda'} + i\,\varepsilon^{\ell
    \alpha\beta\nu} \varepsilon_{\lambda' \alpha} k_{\beta}
  k_{e\nu}\Big],\nonumber\\
\hspace{-0.3in}&&\frac{1}{4}\,{\rm tr}\{(\hat{k}_e + m_e)\,
\vec{\gamma}^{\,i} \hat{k}_{\nu}\gamma^0 \bar{Q}_e(1 - \gamma^5)\} =
E_{\nu} \Big[ (\varepsilon_{\lambda'}\cdot k_e) (2 \vec{k}_e
  + \vec{k})^i - (k_e \cdot k) \vec{\varepsilon}^{\,i}_{\lambda'} +
  i\,\varepsilon^{i \alpha\beta\nu} \varepsilon_{\lambda' \alpha}
  k_{\beta} k_{e\nu}\Big]\nonumber\\
\hspace{-0.3in}&&+ \vec{k}^{\,i}_{\nu} \Big[
  (\varepsilon_{\lambda'}\cdot k_e) (2 E_e + \omega) - (k_e \cdot k)
  \varepsilon^{0}_{\lambda'} + i\,\varepsilon^{0 \alpha\beta\nu}
  \varepsilon_{\lambda' \alpha} k_{\beta} k_{e\nu}\Big]\nonumber\\
\hspace{-0.3in}&&-
i\,\varepsilon^{ij\ell} \vec{k}^{\,j}_{\nu} \Big[
  (\varepsilon_{\lambda'}\cdot k_e) (2 \vec{k}_e + \vec{k})^{\ell} -
  (k_e \cdot k) \vec{\varepsilon}^{\,\ell }_{\lambda'} + i\,\varepsilon^{\ell
    \alpha\beta\nu} \varepsilon_{\lambda' \alpha} k_{\beta}
  k_{e\nu}\Big],\nonumber\\
\hspace{-0.3in}&&\frac{1}{4}\,{\rm tr}\{(\hat{k}_e + m_e)\,
\vec{\gamma}^{\,i} \hat{k}_{\nu}\vec{\gamma}^{\,j}\bar{Q}_e (1
- \gamma^5)\} = \delta^{ij}\,E_{\nu}\Big[
  (\varepsilon_{\lambda'}\cdot k_e) (2 E_e + \omega) - (k_e \cdot k)
  \varepsilon^{0}_{\lambda'} + i\,\varepsilon^{0 \alpha\beta\nu}
  \varepsilon_{\lambda' \alpha} k_{\beta} k_{e\nu}\Big]\nonumber\\
\hspace{-0.3in}&& - i\,\varepsilon^{ij\ell}\,E_{\nu}\Big[
  (\varepsilon_{\lambda'}\cdot k_e) (2 \vec{k}_e + \vec{k})^{\ell} -
  (k_e \cdot k) \vec{\varepsilon}^{\,\ell}_{\lambda'} + i\,\varepsilon^{\ell
    \alpha\beta\nu} \varepsilon_{\lambda' \alpha} k_{\beta}
  k_{e\nu}\Big]\nonumber\\
\hspace{-0.3in}&& - \delta^{ij}\, \vec{k}^{\,\ell}_{\nu} \Big[
  (\varepsilon_{\lambda'}\cdot k_e) (2 \vec{k}_e + \vec{k})^{\ell} -
  (k_e \cdot k) \vec{\varepsilon}^{\,\ell}_{\lambda'} + i\,\varepsilon^{\ell
    \alpha\beta\nu} \varepsilon_{\lambda' \alpha} k_{\beta}
  k_{e\nu}\Big]\nonumber\\
\hspace{-0.3in}&& + \vec{k}^{\,i}_{\nu} \Big[
  (\varepsilon_{\lambda'}\cdot k_e) (2 \vec{k}_e + \vec{k})^j - (k_e
  \cdot k) \varepsilon^{j}_{\lambda'} + i\,\varepsilon^{j
    \alpha\beta\nu} \varepsilon_{\lambda' \alpha} k_{\beta}
  k_{e\nu}\Big]\nonumber\\
\hspace{-0.3in}&& + \vec{k}^{\,j}_{\nu} \Big[
  (\varepsilon_{\lambda'}\cdot k_e) (2 \vec{k}_e + \vec{k})^i - (k_e
  \cdot k) \vec{\varepsilon}^{\,i}_{\lambda'} + i\,\varepsilon^{i
    \alpha\beta\nu} \varepsilon_{\lambda' \alpha} k_{\beta}
  k_{e\nu}\Big] \nonumber\\
\hspace{-0.3in}&& - i\,\varepsilon^{ij\ell}\, \vec{k}^{\,\ell}_{\nu}
\Big[ (\varepsilon_{\lambda'}\cdot k_e) (2 E_e + \omega) - (k_e \cdot
  k) \varepsilon^{0}_{\lambda'} + i\,\varepsilon^{0 \alpha\beta\nu}
  \varepsilon_{\lambda' \alpha} k_{\beta} k_{e\nu}\Big].
\end{eqnarray}
In turn, for the traces without $Q_e$ and $\bar{Q}_e$ we obtain the following expressions
\begin{eqnarray}\label{eq:A.28}
\hspace{-0.3in}\frac{1}{4}\,{\rm tr}\{(\hat{k}_e + m_e)\,
\gamma^0 \hat{k}_{\nu} \gamma^0 (1 - \gamma^5)\} &=& 3 E_e
E_{\nu} + \vec{k}_e \cdot \vec{k}_{\nu},\nonumber\\
\hspace{-0.3in}\frac{1}{4}\,{\rm tr}\{(\hat{k}_e + m_e)\,
\gamma^0\hat{k}_{\nu}\vec{\gamma}^{\,i} (1 - \gamma^5)\} &=& E_e
( \vec{k}_{\nu})^i + E_{\nu}( \vec{k}_e)^i -
i\,\varepsilon^{ij\ell}( \vec{k}_e)^j (
\vec{k}_{\nu})^{\ell},\nonumber\\
\hspace{-0.3in}\frac{1}{4}\,{\rm tr}\{(\hat{k}_e + m_e)\,
\vec{\gamma}^{\,i} \hat{k}_{\nu}\gamma^0 (1 - \gamma^5)\} &=&
E_e ( \vec{k}_{\nu})^i + E_{\nu}( \vec{k}_e)^i +
i\,\varepsilon^{ij\ell}( \vec{k}_e)^j (
\vec{k}_{\nu})^{\ell},\nonumber\\
\hspace{-0.3in}\frac{1}{4}\,{\rm tr}\{(\hat{k}_e + m_e)\,
\vec{\gamma}^{\,i} \hat{k}_{\nu}\vec{\gamma}^{\,j} (1 -
\gamma^5)\} &=& ( \vec{k}_e)^i ( \vec{k}_{\nu})^j + (
\vec{k}_e)^j ( \vec{k}_{\nu})^i + \delta^{ij}(E_e
E_{\nu} - \vec{k}_e \cdot \vec{k}_{\nu})\nonumber\\
\hspace{-0.3in}&+&
i\,\varepsilon^{ij\ell}(E_e( \vec{k}_{\nu})^{\ell}
- E_{\nu}  ( \vec{k}_e)^{\ell}).
\end{eqnarray}
Now we may sum over the photon polarizations. Because of gauge
invariance of Eq.(\ref{eq:A.21}) one may use any gauge.  However, it
is obvious that one has to sum over only physical degrees of freedom
of real photons, which are defined by the polarization vector
$\varepsilon^*_{\lambda'} = (0, \vec{\varepsilon}^{\,*}_{\lambda'})$
\cite{BD1967} (see also \cite{Ivanov2013b}). The polarization vector
$\vec{\varepsilon}^{\,*}_{\lambda'}$ has the following properties
\cite{BD1967} (see also \cite{Ivanov2013b})
\begin{eqnarray}\label{eq:A.29}
\hspace{-0.3in}\vec{k}\cdot \vec{\varepsilon}^{\,*}_{\lambda'} &=&
\vec{k}\cdot \vec{\varepsilon}_{\lambda'} =
0\;,\;\vec{\varepsilon}^{\,*}_{\lambda'}\cdot
\vec{\varepsilon}_{\lambda''} = \delta_{\lambda'\lambda''},\nonumber\\
\hspace{-0.3in}\sum_{\lambda' =
  1,2}\vec{\varepsilon}^{\,i *}_{\lambda'}\vec{\varepsilon}^{\,j}_{\lambda'} &=&
\delta^{ij} - \frac{\vec{k}^{\,i} \vec{k}^{\,j}}{\omega^2} = \delta^{ij} - \vec{n}^{\,i} \vec{n}^{\,j},
\end{eqnarray}
where $\vec{n} = \vec{k}/\omega$. For the summation over only physical
degrees of freedom of a real photon it is convenient to remove from
Eq.(\ref{eq:A.21}) all terms proportional to the time--component of
the photon polarization vector. This gives
\begin{eqnarray*}
\hspace{-0.3in}&&\sum_{\rm pol}\frac{|{\cal M}(n \to p
  e^- \bar{\nu}_e \gamma)_{\lambda'}|^2}{4 m^2_n} = \frac{1}{(k_e\cdot
  k)^2}\,\Big(1 - \frac{E_e + E_{\nu} + \omega}{M}\Big)\,{\rm
  tr}\{(\hat{k}_e + m_e)\,Q_e \gamma^0\hat{k}_{\nu}\gamma^0
\bar{Q}_e (1 - \gamma^5)\}\nonumber\\
\hspace{-0.3in}&& + \frac{1}{(k_e\cdot k)^2}\,\Big[\lambda^2 \Big(1 -
  \frac{E_e + E_{\nu} + \omega}{M}\Big)\,\delta^{ij} -
  \frac{\lambda\,(\kappa + 1)}{M}\,i\,\varepsilon^{ij\ell}(\vec{k}_e +
  \vec{k}_{\nu} + \vec{k}\,)^{\ell}\Big]\,{\rm tr}\{(\hat{k}_e
+ m_e)\,Q_e \vec{\gamma}^{\,i}\, \hat{k}_{\nu}
\vec{\gamma}^{\,j}\, \bar{Q}_e (1 - \gamma^5)\} \nonumber\\
\hspace{-0.3in}&& + \frac{1}{(k_e\cdot k)^2}\,\frac{\lambda^2 +
  1}{2M}\,(\vec{k}_e + \vec{k}_{\nu} + \vec{k}\,) \cdot{\rm tr}\{(\hat{k}_e +
m_e)\,Q_e \gamma^0\hat{k}_{\nu}\vec{\gamma}\, \bar{Q}_e (1 -
\gamma^5)\}\nonumber\\
\hspace{-0.3in}&& + \frac{1}{(k_e\cdot k)^2}\,\frac{\lambda^2 +
  1}{2M}\,(\vec{k}_e + \vec{k}_{\nu} + \vec{k}\,) \cdot {\rm
  tr}\{(\hat{k}_e + m_e)\,Q_e \vec{\gamma}\,\hat{k}_{\nu}
\gamma^0 \bar{Q}_e (1 - \gamma^5)\}\nonumber\\
\hspace{-0.3in}&& +  \frac{1}{k_e\cdot k}\,\Big(- \frac{\lambda}{M}\,
\vec{\varepsilon}^{\,*}_{\lambda'} + \frac{\lambda}{M
  \omega}\,i\,(\vec{\varepsilon}^{\,*}_{\lambda'} \times
\vec{k}\,)\Big)\cdot {\rm tr}\{(\hat{k}_e +
m_e)\,\vec{\gamma}\,\hat{k}_{\nu}\gamma^0 \, \bar{Q}_e (1 -
\gamma^5)\}\nonumber\\
\hspace{-0.3in}&& + \frac{1}{k_e\cdot k}\,\Big(- \frac{\lambda}{M }\,
\vec{\varepsilon}_{\lambda'} - \frac{\lambda}{M
  \omega}\,i\,(\vec{\varepsilon}_{\lambda'} \times
\vec{k}\,)\Big)\cdot {\rm tr}\{(\hat{k}_e + m_e)\,Q_e
\gamma^0\hat{k}_{\nu}\vec{\gamma}\,(1 - \gamma^5)\}\nonumber\\
\end{eqnarray*}
\begin{eqnarray}\label{eq:A.30}
\hspace{-0.3in}&& + \frac{1}{k_e\cdot k}\,\Big(- \frac{\lambda^2}{M}\,
\vec{\varepsilon}^{\,*}_{\lambda'} + i\,\frac{\lambda}{M
  \omega}\,(\vec{\varepsilon}^{\,*}_{\lambda'} \times
\vec{k}\,)\Big)\cdot {\rm tr}\{(\hat{k}_e + m_e)\,\gamma^0
\,\hat{k}_{\nu} \vec{\gamma}\,\bar{Q}_e (1 -
\gamma^5)\}\nonumber\\
\hspace{-0.3in}&& + \frac{1}{k_e\cdot k}\,\Big(-\frac{\lambda^2}{M}\,
\vec{\varepsilon}_{\lambda'} - \frac{\lambda}{M
  \omega}\,i\,(\vec{\varepsilon}_{\lambda'} \times \vec{k}\,) \Big)
\cdot {\rm tr}\{(\hat{k}_e + m_e)\,Q_e
\vec{\gamma}\,\hat{k}_{\nu}\gamma^0 (1 - \gamma^5)\}\nonumber\\
\hspace{-0.3in}&& + \frac{1}{k_e\cdot k}\,\Big[\frac{\lambda^2 }{ M
    \omega} \Big(\vec{\varepsilon}^{\,i *}_{\lambda'}\vec{k}^{\,j} -
  \vec{\varepsilon}^{\,j *}_{\lambda'} \vec{k}^{\,i}\Big) +
  \frac{\lambda^2}{M }\,i\,\varepsilon^{ij\ell}
  \vec{\varepsilon}^{\,\ell *}_{\lambda'}\Big)\Big] {\rm
  tr}\{(\hat{k}_e + m_e)\, \vec{\gamma}^{\,i}\,\hat{k}_{\nu}
\vec{\gamma}^{\,j}\,\bar{Q}_e (1 - \gamma^5)\}\nonumber\\
\hspace{-0.3in}&& - \frac{1}{k_e\cdot k}\,\Big[\frac{\lambda^2}{M
    \omega}\,\Big(\vec{\varepsilon}^{\,i}_{\lambda'}\vec{k}^{\,j} -
  \vec{\varepsilon}^{\,j}_{\lambda'} \vec{k}^{\,i}\Big)
  -\frac{\lambda^2}{M }\,i\,\varepsilon^{ij\ell}\,
  \vec{\varepsilon}^{\,\ell}_{\lambda'}\Big]\,{\rm tr}\{(\hat{k}_e +
m_e)\,Q_e \vec{\gamma}^{\,i}\,\hat{k}_{\nu}
\vec{\gamma}^{\,j}\,(1 - \gamma^5)\}.
\end{eqnarray}
Summing over the photon polarizations we obtain
\begin{eqnarray*}
\hspace{-0.3in}&&\frac{1}{16}\sum_{\lambda'}{\rm tr}\{(\hat{k}_e +
m_e)\,Q_e \gamma^0\hat{k}_{\nu}\gamma^0 \bar{Q}_e (1 - \gamma^5)\} =
E_eE_{\nu}\Big\{\Big[\Big(1 + \frac{\omega}{E_e}\Big)\Big(k^2_e -
  (\vec{k}_e\cdot \vec{n}\,)^2\Big) + \frac{\omega^2}{E_e}\Big(E_e -
  \vec{k}_e \cdot \vec{n}\,\Big)\Big]\nonumber\\
\hspace{-0.3in}&& + \frac{\vec{k}_e\cdot \vec{k}_{\nu}}{E_e
  E_{\nu}}\Big[\Big(k^2_e - ( \vec{k}_e\cdot \vec{n}\,)^2\Big) +
  \omega \Big(E_e - \vec{k}_e\cdot \vec{n}\,\Big)\Big] +
\frac{\vec{k}_{\nu}\cdot \vec{n}}{ E_{\nu}}\Big[\omega \Big(1 +
  \frac{\omega}{E_e}\Big)\Big(E_e - \vec{k}_e\cdot \vec{n}\,\Big) -
  \omega \frac{m^2_e}{E_e}\Big]\Big\},\nonumber\\
\hspace{-0.3in}&& \frac{1}{16}\sum_{\lambda'}{\rm tr}\{(\hat{k}_e +
m_e)\,Q_e \vec{\gamma}^{\,i}\, \hat{k}_{\nu} \vec{\gamma}^{\,j}\,
\bar{Q}_e (1 - \gamma^5)\} = E_{\nu}E_e \Big\{\delta^{ij}\Big[\Big(1 +
  \frac{\omega}{E_e}\Big)\Big(k^2_e - (\vec{k}_e\cdot
  \vec{n}\,)^2\Big) + \frac{\omega^2}{E_e}\Big(E_e - \vec{k}_e \cdot
  \vec{n}\,\Big)\Big]\nonumber\\
\hspace{-0.3in}&& - \delta^{ij}\Big\{\frac{\vec{k}_e\cdot
  \vec{k}_{\nu}}{E_e E_{\nu}}\Big[\Big(k^2_e - ( \vec{k}_e\cdot
  \vec{n}\,)^2\Big) + \omega \Big(E_e - \vec{k}_e\cdot
  \vec{n}\,\Big)\Big] + \frac{\vec{k}_{\nu}\cdot \vec{n}}{
  E_{\nu}}\Big[\omega \Big(1 + \frac{\omega}{E_e}\Big)\Big(E_e -
  \vec{k}_e\cdot \vec{n}\,\Big) - \omega \frac{m^2_e}{E_e}\Big\}
  \nonumber\\
\hspace{-0.3in}&& - \,i\,\varepsilon^{ij\ell}\Big[\Big(k^2_e -
  (\vec{k}_e\cdot \vec{n}\,)^2\Big)\,\frac{(\vec{k}_e + \omega
    \vec{n}\,)^{\ell}}{E_e} + \frac{\omega}{E_e}
  \Big(\vec{k}^{\,\ell}_e - \vec{n}^{\,\ell} (\vec{k}_e \cdot
  \vec{n}\,)\Big) \Big(E_e - \vec{k}_e\cdot \vec{n}\,\Big) +
  \frac{\omega^2}{E_e} \,\vec{n}^{\,\ell}\,\Big(E_e - \vec{k}_e\cdot
  \vec{n}\,\Big)\Big]\nonumber\\
\hspace{-0.3in}&& - \,i\,\varepsilon^{ij\ell}
\frac{\vec{k}^{\,\ell}_{\nu}}{E_{\nu}}\Big[\Big(1 +
  \frac{\omega}{E_e}\Big)\Big(k^2_e - (\vec{k}_e\cdot
  \vec{n}\,)^2\Big) + \frac{\omega^2}{E_e}\Big(E_e - \vec{k}_e \cdot
  \vec{n}\,\Big)\Big] +
\frac{\vec{k}^{\,i}_{\nu}}{E_{\nu}}\Big[\Big(k^2_e - (\vec{k}_e\cdot
  \vec{n}\,)^2\Big)\,\frac{(\vec{k}_e + \omega \vec{n}\,)^j}{E_e}
  \nonumber\\
\hspace{-0.3in}&& + \frac{\omega}{E_e} \Big(\vec{k}^{\,j}_e -
\vec{n}^{\,j} (\vec{k}_e \cdot \vec{n}\,)\Big) \Big(E_e -
\vec{k}_e\cdot \vec{n}\,\Big) + \frac{\omega^2}{E_e}\,\vec{n}^{\,j}
\Big(E_e - \vec{k}_e\cdot \vec{n}\,\Big) \Big] +
\frac{\vec{k}^{\,j}_{\nu}}{E_{\nu}}\Big[\Big(k^2_e - (\vec{k}_e\cdot
  \vec{n}\,)^2\Big)\,\frac{(\vec{k}_e + \omega \vec{n}\,)^i}{E_e}
  \nonumber\\
\hspace{-0.3in}&& + \frac{\omega}{E_e} \Big(\vec{k}^{\,i}_e -
\vec{n}^{\,i} (\vec{k}_e \cdot \vec{n}\,)\Big) \Big(E_e -
\vec{k}_e\cdot \vec{n}\,\Big) + \frac{\omega^2}{E_e}\,\vec{n}^{\,i}
\Big(E_e - \vec{k}_e\cdot \vec{n}\,\Big) \Big]\Big\} \nonumber\\
\hspace{-0.3in}&&\frac{1}{16}\sum_{\lambda'}{\rm tr}\{(\hat{k}_e +
m_e)\,Q_e \vec{\gamma}\, \hat{k}_{\nu}\,\gamma^0 \,\bar{Q}_e (1 -
\gamma^5)\} = E_{\nu}E_e\Big\{\Big[\Big(k^2_e - (\vec{k}_e\cdot
  \vec{n}\,)^2\Big)\,\frac{(\vec{k}_e + \omega \vec{n}\,)}{E_e}+
  \frac{\omega}{E_e} \Big(\vec{k}_e - \vec{n}\, (\vec{k}_e \cdot
  \vec{n}\,)\Big)\nonumber\\
\hspace{-0.3in}&&\times \Big(E_e - \vec{k}_e\cdot \vec{n}\,\Big) + \vec{n}\,\frac{\omega^2}{E_e}\Big(E_e - \vec{k}_e
\cdot \vec{n}\,\Big)\Big] + \frac{\vec{k}_{\nu}}{E_{\nu}}\Big[\Big(1 +
  \frac{\omega}{E_e}\Big)\Big(k^2_e - (\vec{k}_e\cdot
  \vec{n}\,)^2\Big) + \frac{\omega^2}{E_e}\Big(E_e - \vec{k}_e \cdot
  \vec{n}\,\Big)\Big]\nonumber\\
\hspace{-0.3in}&& + i\Big[\Big(k^2_e - (\vec{k}_e\cdot
  \vec{n}\,)^2\Big)\frac{\vec{k}_{\nu}\times (\vec{k}_e + \omega
    \vec{n}\,)}{E_{\nu}E_e} +
  \frac{\omega^2}{E_e}\frac{\vec{k}_{\nu}\times \vec{n}}{E_{\nu}}
  \Big(E_e - \vec{k}_e\cdot \vec{n}\,\Big) + \omega
  \Big(\frac{\vec{k}_{\nu}\times \vec{k}_e}{E_{\nu}E_e} -
  \frac{(\vec{k}_{\nu}\times \vec{n}\,)(\vec{k}_e\cdot
    \vec{n}\,)}{E_{\nu}E_e}\Big)\Big(E_e - \vec{k}_e\cdot
  \vec{n}\,\Big)\Big\},\nonumber\\
\hspace{-0.3in}&&\frac{1}{16}\sum_{\lambda'}{\rm tr}\{(\hat{k}_e +
m_e)\,Q_e \gamma^0 \,\hat{k}_{\nu}\,\vec{\gamma}\, \bar{Q}_e (1 -
\gamma^5)\} = E_{\nu}E_e\Big\{\Big[\Big(k^2_e - (\vec{k}_e\cdot
  \vec{n}\,)^2\Big)\,\frac{(\vec{k}_e + \omega \vec{n}\,)}{E_e}+
  \frac{\omega}{E_e} \Big(\vec{k}_e - \vec{n}\, (\vec{k}_e \cdot
  \vec{n}\,)\Big)\nonumber\\
\hspace{-0.3in}&&\times\, \Big(E_e - \vec{k}_e\cdot \vec{n}\,\Big) +
\vec{n}\,\frac{\omega^2}{E_e}\Big(E_e - \vec{k}_e \cdot
\vec{n}\,\Big)\Big] + \frac{\vec{k}_{\nu}}{E_{\nu}}\Big[\Big(1 +
  \frac{\omega}{E_e}\Big)\Big(k^2_e - (\vec{k}_e\cdot
  \vec{n}\,)^2\Big) + \frac{\omega^2}{E_e}\Big(E_e - \vec{k}_e \cdot
  \vec{n}\,\Big)\Big]\nonumber\\
\hspace{-0.3in}&& - i\Big[\Big(k^2_e - (\vec{k}_e\cdot
  \vec{n}\,)^2\Big)\frac{\vec{k}_{\nu}\times (\vec{k}_e + \omega
    \vec{n}\,)}{E_{\nu}E_e} +
  \frac{\omega^2}{E_e}\frac{\vec{k}_{\nu}\times \vec{n}}{E_{\nu}}
  \Big(E_e - \vec{k}_e\cdot \vec{n}\,\Big) + \omega
  \Big(\frac{\vec{k}_{\nu}\times \vec{k}_e}{E_{\nu}E_e} -
  \frac{(\vec{k}_{\nu}\times \vec{n}\,)(\vec{k}_e\cdot
    \vec{n}\,)}{E_{\nu}E_e}\Big)\Big(E_e - \vec{k}_e\cdot
  \vec{n}\,\Big)\Big\},\nonumber\\
\hspace{-0.3in}&& \frac{1}{4}
\sum_{\lambda'}\vec{\varepsilon}^{\,i*}_{\lambda'} {\rm
  tr}\{(\hat{k}_e + m_e)\,\vec{\gamma}^{\,j}\,\hat{k}_{\nu}\gamma^0 \,
\bar{Q}_e (1 - \gamma^5)\} = E_{\nu}\Big[ -\Big(\vec{k}^{\,i}_e -
  \vec{n}^{\,i}(\vec{k}_e\cdot \vec{n}\,)\Big)(2 \vec{k}_e +
 \omega \vec{n}\,)^j - \Big(\delta^{ij} - \vec{n}^{\,i}
  \vec{n}^{\,j}\Big)\nonumber\\
\hspace{-0.3in}&&\times\,\omega\,\Big(E_e - \vec{k}_e\cdot
\vec{n}\,\Big) - \omega\,\vec{n}^{\,i}\,i\,\varepsilon^{j\ell
  a}\,\vec{n}^{\,\ell}\,\vec{k}^{\,a}_e -
\omega\,i\,\varepsilon^{ij\ell}\,\vec{k}^{\,\ell}_e +
\omega\,E_e\,i\,\varepsilon^{ij\ell}\,\vec{n}^{\,\ell}\Big] +
\vec{k}^{\,j}_{\nu} \Big[-\Big(\vec{k}^{\,i}_e -
  \vec{n}^{\,i}(\vec{k}_e\cdot \vec{n}\,)\Big)\nonumber\\ 
\hspace{-0.3in}&&\times\,(2 E_e + \omega) -
\omega\,i\,\varepsilon^{i\ell a}\,\vec{n}^{\,\ell}\,\vec{k}^{\,a}_e\Big] -
i\,\varepsilon^{ja\ell}\,\vec{k}^{\,a}_{\nu} \Big[-\Big(\vec{k}^{\,i}_e -
  \vec{n}^{\,i}(\vec{k}_e\cdot \vec{n}\,)\Big)(2 \vec{k}_e +
  \omega  \vec{n}\,)^{\ell} - \Big(\delta^{i\ell} -
  \vec{n}^{\,i}\,\vec{n}^{\,\ell}\Big)\nonumber\\ 
\hspace{-0.3in}&&\times\,\omega\,\Big(E_e - \vec{k}_e\cdot
\vec{n}\,\Big) - \omega\,\vec{n}^{\,i}\,i\,\varepsilon^{\ell b
  c}\,\vec{n}^{\,b}\,\vec{k}^{\,c}_e  + \omega\,i\,\varepsilon^{\ell i
  b}\vec{k}^{\,b}_e - \omega\,E_e\,i\,\varepsilon^{\ell i
  a}\,\vec{n}^{\,a}\Big],\nonumber\\
\hspace{-0.3in}&&\frac{1}{4}
\sum_{\lambda'}\vec{\varepsilon}^{\,i}_{\lambda'} {\rm tr}\{(\hat{k}_e
+ m_e)\,Q_e \gamma^0\,\hat{k}_{\nu}\vec{\gamma}^{\,j}\, (1 -
\gamma^5)\} = E_{\nu}\Big[ -\Big(\vec{k}^{\,i}_e -
  \vec{n}^{\,i}(\vec{k}_e\cdot \vec{n}\,)\Big)(2 \vec{k}_e + \omega
  \vec{n}\,)^j - \Big(\delta^{ij} - \vec{n}^{\,i}
  \vec{n}^{\,j}\Big)\nonumber\\
 \end{eqnarray*}
\begin{eqnarray}\label{eq:A.31}
\hspace{-0.3in}&&\times\,\omega\,\Big(E_e - \vec{k}_e\cdot
\vec{n}\,\Big) + \omega\,\vec{n}^{\,i}\,i\,\varepsilon^{j\ell
  a}\,\vec{n}^{\,\ell}\,\vec{k}^{\,a}_e +
\omega\,i\,\varepsilon^{ij\ell}\,\vec{k}^{\,\ell}_e -
\omega\,E_e\,i\,\varepsilon^{ij\ell}\,\vec{n}^{\,\ell}\Big] +
\vec{k}^{\,j}_{\nu} \Big[-\Big(\vec{k}^{\,i}_e -
  \vec{n}^{\,i}(\vec{k}_e\cdot \vec{n}\,)\Big)\nonumber\\ 
\hspace{-0.3in}&&\times\, (2 E_e + \omega) +
\omega\,i\,\varepsilon^{i\ell a}\,\vec{n}^{\,\ell}\,\vec{k}^{\,a}_e\Big] +
i\,\varepsilon^{ja\ell}\,\vec{k}^{\,a}_{\nu} \Big[-\Big(\vec{k}^{\,i}_e -
  \vec{n}^{\,i}(\vec{k}_e\cdot \vec{n}\,)\Big)(2 \vec{k}_e +
   \omega \vec{n}\,)^{\ell} - \Big(\delta^{i\ell} -
  \vec{n}^{\,i}\,\vec{n}^{\,\ell}\Big)\nonumber\\ 
\hspace{-0.3in}&&\times\,\omega\,\Big(E_e - \vec{k}_e\cdot
\vec{n}\,\Big) + \omega\,\vec{n}^{\,i}\,i\,\varepsilon^{\ell b
  c}\,\vec{n}^{\,b}\,\vec{k}^{\,c}_e - \omega\,i\,\varepsilon^{\ell i
  b}\,\vec{k}^{\,b}_e + \omega\,E_e\,i\,\varepsilon^{\ell i
  a}\,\vec{n}^{\,a}\Big],\nonumber\\
\hspace{-0.3in}&& \frac{1}{4} \sum_{\lambda'}\vec{\varepsilon}^{\,
  i*}_{\lambda'} {\rm tr}\{(\hat{k}_e + m_e)\,\gamma^0 \,\hat{k}_{\nu}
\vec{\gamma}^{\,j}\,\bar{Q}_e (1 - \gamma^5)\} = E_{\nu} \Big[ -
  \Big(\vec{k}^{\,i}_e - \vec{n}^{\,i}(\vec{k}_e\cdot
  \vec{n}\,)\Big)(2 \vec{k}_e +  \omega \vec{n}\,)^j - \Big(\delta^{ij} -
  \vec{n}^{\,i} \vec{n}^{\,j}\Big)\nonumber\\
\hspace{-0.3in}&&\times\,\omega\,\Big(E_e - \vec{k}_e\cdot
\vec{n}\,\Big) - \omega\,\vec{n}^{\,i}\,i\,\varepsilon^{j\ell
  a}\,\vec{n}^{\,\ell}\,\vec{k}^{\,a}_e -
\omega\,i\,\varepsilon^{ij\ell}\,\vec{k}^{\,\ell}_e +
\omega\,E_e\,i\,\varepsilon^{ij\ell}\,\vec{n}^{\,\ell}\Big] +
\vec{k}^{\,j}_{\nu} \Big[- \Big(\vec{k}^{\,i}_e -
  \vec{n}^{\,i}(\vec{k}_e\cdot \vec{n}\,)\Big) (2 E_e + \omega)\nonumber\\
\hspace{-0.3in}&& - \omega\,i\,\varepsilon^{j\ell
  a}\,\vec{n}^{\,\ell}\,\vec{k}^{\,a}_e\Big] -
i\,\varepsilon^{j a \ell}\vec{k}^{\,a}_{\nu}\Big[ - \Big(\vec{k}^{\,i}_e
  - \vec{n}^{\,i}(\vec{k}_e\cdot \vec{n}\,)\Big)(2 \vec{k}_e +
  \omega  \vec{n}\,)^{\ell} - \Big(\delta^{i\ell} -
  \vec{n}^{\,i} \vec{n}^{\,\ell}\Big)\,\omega\,\Big(E_e - \vec{k}_e\cdot
\vec{n}\,\Big) - \omega\,\vec{n}^{\,i}\,i\,\varepsilon^{\ell a
  b}\,\vec{n}^{\,a}\,\vec{k}^{\,b}_e\nonumber\\
\hspace{-0.3in}&& + \omega\,i\,\varepsilon^{\ell i b}\,\vec{k}^{\,b}_e
- \omega\,E_e\,i\,\varepsilon^{\ell i b}\,\vec{n}^{\,b}\Big],\nonumber\\
\hspace{-0.3in}&&\frac{1}{4} \sum_{\lambda'}\vec{\varepsilon}^{\,
  i}_{\lambda'} {\rm tr}\{(\hat{k}_e +
m_e)\,Q_e\,\vec{\gamma}^{\,j}\,\hat{k}_{\nu} \gamma^0\, (1 -
\gamma^5)\} =  E_{\nu}\Big[ -
  \Big(\vec{k}^{\,i}_e - \vec{n}^{\,i}(\vec{k}_e\cdot
  \vec{n}\,)\Big)(2 \vec{k}_e +  \omega \vec{n}\,)^j - \Big(\delta^{ij} -
  \vec{n}^{\,i} \vec{n}^{\,j}\Big)\nonumber\\
\hspace{-0.3in}&&\times\,\omega\,\Big(E_e - \vec{k}_e\cdot
\vec{n}\,\Big) + \omega\,\vec{n}^{\,i}\,i\,\varepsilon^{j\ell
  a}\,\vec{n}^{\,\ell}\,\vec{k}^{\,a}_e +
\omega\,i\,\varepsilon^{ij\ell}\,\vec{k}^{\,\ell}_e  -
\omega\,E_e\,i\,\varepsilon^{ij\ell}\,\vec{n}^{\,\ell}\Big] + 
\vec{k}^{\,j}_{\nu} \Big[- \Big(\vec{k}^{\,i}_e -
  \vec{n}^{\,i}(\vec{k}_e\cdot \vec{n}\,)\Big) (2 E_e + \omega)\nonumber\\
\hspace{-0.3in}&& + \omega\,i\,\varepsilon^{i \ell
  a}\,\vec{n}^{\,\ell}\,\vec{k}^{\,a}_e\Big] - i\,\varepsilon^{j a
  \ell}\vec{k}^{\,a}_{\nu}\Big[ - \Big(\vec{k}^{\,i}_e -
  \vec{n}^{\,i}(\vec{k}_e\cdot \vec{n}\,)\Big)(2 \vec{k}_e + \omega
  \vec{n}\,)^{\ell} - \Big(\delta^{i\ell} - \vec{n}^{\,i}
  \vec{n}^{\,\ell}\Big)\,\omega\,\Big(E_e - \vec{k}_e\cdot
  \vec{n}\,\Big) + \omega\,\vec{n}^{\,i}\,i\,\varepsilon^{\ell a
    b}\,\vec{n}^{\,a}\,\vec{k}^{\,b}_e\nonumber\\
\hspace{-0.3in}&& - \omega\,i\,\varepsilon^{\ell i b}\,\vec{k}^{\,b}_e
+ \omega\,E_e\,i\,\varepsilon^{\ell i b}\,\vec{n}^{\,b}\Big],\nonumber\\
\hspace{-0.3in}&&\frac{1}{4}\sum_{\lambda'}\vec{\varepsilon}^{\,\ell
  *}_{\lambda' } {\rm tr}\{(\hat{k}_e + m_e)\,
\vec{\gamma}^{\,i}\,\hat{k}_{\nu} \vec{\gamma}^{\,j}\,\bar{Q}_e (1 -
\gamma^5)\} = \delta^{ij} E_{\nu}\Big[ - \Big(\vec{k}^{\,\ell}_e -
  \vec{n}^{\,\ell}(\vec{k}_e\cdot \vec{n}\,)\Big)(2 E_e + \omega) -
  \omega\,i\,\varepsilon^{\ell b c} \vec{n}^{\,b} \vec{k}^{\,c}_e\Big]
\nonumber\\
\hspace{-0.3in}&& - i\,\varepsilon^{ija} E_{\nu}\Big[-
  \Big(\vec{k}^{\,\ell}_e - \vec{n}^{\,\ell}(\vec{k}_e\cdot
  \vec{n}\,)\Big)(2 \vec{k}_e + \omega \vec{n}\,)^{\,a} - \omega
  \Big(\delta^{\ell a} - \vec{n}^{\,\ell} \vec{n}^{\,a}\Big) \Big(E_e
  - \vec{k}_e \cdot \vec{n}\,\Big) -
  \omega\,\vec{n}^{\,\ell}\,i\,\varepsilon^{abc}
  \vec{n}^{\,b}\vec{k}^{\,c}_e + \omega\,i\,\varepsilon^{a\ell b}
  \vec{k}^{\,b}_e\nonumber\\
\hspace{-0.3in}&& - \omega E_e\,i\,\varepsilon^{a\ell b} \vec{n}^{\,
  b}\Big] - \delta^{ij} \vec{k}^{\,a}_{\nu}\Big[-
  \Big(\vec{k}^{\,\ell}_e - \vec{n}^{\,\ell}(\vec{k}_e\cdot
  \vec{n}\,)\Big)(2 \vec{k}_e + \omega \vec{n}\,)^{\,a} - \omega
  \Big(\delta^{\ell a} - \vec{n}^{\,\ell} \vec{n}^{\,a}\Big) \Big(E_e
  - \vec{k}_e \cdot \vec{n}\,\Big) -
  \omega\,\vec{n}^{\,\ell}\,i\,\varepsilon^{abc}
  \vec{n}^{\,b}\vec{k}^{\,c}_e \nonumber\\
\hspace{-0.3in}&& + \omega\,i\,\varepsilon^{a\ell b} \vec{k}^{\,b}_e - 
\omega E_e\,i\,\varepsilon^{a\ell b} \vec{n}^{\, b}\Big] +
\vec{k}^{\,i}_{\nu}\Big[- \Big(\vec{k}^{\,\ell}_e -
  \vec{n}^{\,\ell}(\vec{k}_e\cdot \vec{n}\,)\Big)(2 \vec{k}_e + \omega
  \vec{n}\,)^{\,j} - \omega \Big(\delta^{\ell j} - \vec{n}^{\,\ell}
  \vec{n}^{\,j}\Big) \Big(E_e - \vec{k}_e \cdot \vec{n}\,\Big) \nonumber\\
\hspace{-0.3in}&& - \omega\,\vec{n}^{\,\ell}\,i\,\varepsilon^{jbc}
\vec{n}^{\,b}\vec{k}^{\,c}_e + \omega\,i\,\varepsilon^{j\ell b}
\vec{k}^{\,b}_e - \omega E_e\,i\,\varepsilon^{j\ell b} \vec{n}^{\,
  b}\Big] +  \vec{k}^{\,j}_{\nu}\Big[- \Big(\vec{k}^{\,\ell}_e -
  \vec{n}^{\,\ell}(\vec{k}_e\cdot \vec{n}\,)\Big)(2 \vec{k}_e + \omega
  \vec{n}\,)^{\,i} - \omega \Big(\delta^{\ell i} - \vec{n}^{\,\ell}
  \vec{n}^{\,i}\Big) \Big(E_e - \vec{k}_e \cdot \vec{n}\,\Big)
  \nonumber\\
\hspace{-0.3in}&& - \omega\,\vec{n}^{\,\ell}\,i\,\varepsilon^{ibc}
\vec{n}^{\,b}\vec{k}^{\,c}_e + \omega\,i\,\varepsilon^{i\ell b}
\vec{k}^{\,b}_e  - \omega E_e\,i\,\varepsilon^{i\ell b} \vec{n}^{\,
  b}\Big] - i\,\varepsilon^{ija} \vec{k}^{\,a}_{\nu}\Big[ -
  \Big(\vec{k}^{\,\ell}_e - \vec{n}^{\,\ell}(\vec{k}_e\cdot
  \vec{n}\,)\Big)(2 E_e + \omega) - \omega\,i\,\varepsilon^{\ell b c}
  \vec{n}^{\,b} \vec{k}^{\,c}_e\Big],\nonumber\\
\hspace{-0.3in}&&\frac{1}{4}\sum_{\lambda'}\vec{\varepsilon}^{\,\ell
}_{\lambda' } {\rm tr}\{(\hat{k}_e + m_e)\,
Q_e\,\vec{\gamma}^{\,i}\,\hat{k}_{\nu} \vec{\gamma}^{\,j}\,(1 -
\gamma^5)\} = \delta^{ij} E_{\nu}\Big[ - \Big(\vec{k}^{\,\ell}_e -
  \vec{n}^{\,\ell}(\vec{k}_e\cdot \vec{n}\,)\Big)(2 E_e + \omega) +
  \omega\,i\,\varepsilon^{\ell b c} \vec{n}^{\,b} \vec{k}^{\,c}_e\Big]
\nonumber\\
\hspace{-0.3in}&& - i\,\varepsilon^{ija} E_{\nu}\Big[-
  \Big(\vec{k}^{\,\ell}_e - \vec{n}^{\,\ell}(\vec{k}_e\cdot
  \vec{n}\,)\Big)(2 \vec{k}_e + \omega \vec{n}\,)^{\,a} - \omega
  \Big(\delta^{\ell a} - \vec{n}^{\,\ell} \vec{n}^{\,a}\Big) \Big(E_e
  - \vec{k}_e \cdot \vec{n}\,\Big) +
  \omega\,\vec{n}^{\,\ell}\,i\,\varepsilon^{abc}
  \vec{n}^{\,b}\vec{k}^{\,c}_e - \omega\,i\,\varepsilon^{a\ell b}
  \vec{k}^{\,b}_e\nonumber\\
\hspace{-0.3in}&& + \omega E_e\,i\,\varepsilon^{a\ell b} \vec{n}^{\,
  b}\Big] - \delta^{ij} \vec{k}^{\,a}_{\nu}\Big[-
  \Big(\vec{k}^{\,\ell}_e - \vec{n}^{\,\ell}(\vec{k}_e\cdot
  \vec{n}\,)\Big)(2 \vec{k}_e + \omega \vec{n}\,)^{\,a} - \omega
  \Big(\delta^{\ell a} - \vec{n}^{\,\ell} \vec{n}^{\,a}\Big) \Big(E_e
  - \vec{k}_e \cdot \vec{n}\,\Big)  +
  \omega\,\vec{n}^{\,\ell}\,i\,\varepsilon^{abc}
  \vec{n}^{\,b}\vec{k}^{\,c}_e \nonumber\\
\hspace{-0.3in}&& - \omega\,i\,\varepsilon^{a\ell b} \vec{k}^{\,b}_e +
\omega E_e\,i\,\varepsilon^{a\ell b} \vec{n}^{\, b}\Big] +
\vec{k}^{\,i}_{\nu}\Big[- \Big(\vec{k}^{\,\ell}_e -
  \vec{n}^{\,\ell}(\vec{k}_e\cdot \vec{n}\,)\Big)(2 \vec{k}_e + \omega
  \vec{n}\,)^{\,j} - \omega \Big(\delta^{\ell j} - \vec{n}^{\,\ell}
  \vec{n}^{\,j}\Big) \Big(E_e - \vec{k}_e \cdot \vec{n}\,\Big) \nonumber\\
\hspace{-0.3in}&& + \omega\,\vec{n}^{\,\ell}\,i\,\varepsilon^{jbc}
\vec{n}^{\,b}\vec{k}^{\,c}_e - \omega\,i\,\varepsilon^{j\ell b}
\vec{k}^{\,b}_e + \omega E_e\,i\,\varepsilon^{j\ell b} \vec{n}^{\,
  b}\Big] + 
\vec{k}^{\,j}_{\nu}\Big[- \Big(\vec{k}^{\,\ell}_e -
  \vec{n}^{\,\ell}(\vec{k}_e\cdot \vec{n}\,)\Big)(2 \vec{k}_e + \omega
  \vec{n}\,)^{\,i} - \omega \Big(\delta^{\ell i} - \vec{n}^{\,\ell}
  \vec{n}^{\,i}\Big) \Big(E_e - \vec{k}_e \cdot \vec{n}\,\Big) \nonumber\\
\hspace{-0.3in}&& +\omega\,\vec{n}^{\,\ell}\,i\,\varepsilon^{ibc}
\vec{n}^{\,b}\vec{k}^{\,c}_e - \omega\,i\,\varepsilon^{i\ell b}
\vec{k}^{\,b}_e + \omega E_e\,i\,\varepsilon^{i\ell b} \vec{n}^{\,
  b}\Big] - i\,\varepsilon^{ija} \vec{k}^{\,a}_{\nu}\Big[ -
  \Big(\vec{k}^{\,\ell}_e - \vec{n}^{\,\ell}(\vec{k}_e\cdot
  \vec{n}\,)\Big)(2 E_e + \omega) + \omega\,i\,\varepsilon^{\ell b c}
  \vec{n}^{\,b} \vec{k}^{\,c}_e\Big].
\end{eqnarray}
The rate of the neutron radiative $\beta^-$--decay is given by
 \begin{eqnarray}\label{eq:A.32}
\lambda_{\beta \gamma} &=& \pi \alpha G^2_F |V_{ud}|^2
\frac{1}{m_n}\int \frac{1}{2 }\sum_{\rm \lambda', pol}|{\cal M}(n \to
p e^- \bar{\nu}_e\gamma)_{\lambda'}|^2 F(E_e, Z =
1)\nonumber\\ &&\times\, (2\pi)^4\delta^{(4)}(k_n - k_p - k_e -
k_{\nu} - k)\,\frac{d^3k_p}{(2\pi)^3 2 E_p}\frac{d^3k_e}{(2\pi)^3 2
  E_e}\frac{d^3k_{\nu}}{(2\pi)^3 2 E_{\nu}}\frac{d^3k}{(2\pi)^3 2
  \omega},
\end{eqnarray}
where $F(E_e, Z = 1)$ is the relativistic Fermi function, taking into
account the final--state proton--electron Coulomb interaction
\cite{Ivanov2013,Ivanov2013a} (see also Eq.(\ref{eq:8})). After the
integration over the proton momentum $\vec{k}_p$, the directions of
the momenta of the particles in the final state we are left with the
result \cite{Ivanov2013}
\begin{eqnarray}\label{eq:A.33}
\hspace{-0.3in}&&\lambda_{\beta \gamma}(\omega_{\rm max}, \omega_{\rm
  min}) = (1 + 3 \lambda^2)\frac{\alpha}{\pi} \frac{G^2_F
  |V_{ud}|^2}{2\pi^3} \int^{\omega_{\rm max}}_{\omega_{\rm
    min}}\!\!\frac{d\omega}{\omega}\int^{E_0 - \omega}_{m_e}\!\!\!\!\!\!\!
dE_e\,E_e\sqrt{E^2_e - m^2_e} (E_0 - E_e - \omega)^2\,F(E_e, Z =
1)\,\rho_{\beta\gamma}(E_e,\omega),\nonumber\\
\hspace{-0.3in}&&
\end{eqnarray}
where the rate of the neutron radiative $\beta^-$--decay is defined
for the photon energy region $\omega_{\rm min} \le \omega \le
\omega_{\rm max}$ and $E_0 = (m^2_n - m^2_p + m^2_e)/2 m_n$ is the
end--point energy of the electron energy--spectrum of the neutron
$\beta^-$--decay \cite{Ivanov2013}. The function
$\rho_{\beta\gamma}(E_e, \omega)$ is given by
\begin{eqnarray}\label{eq:A.34}
\rho_{\beta\gamma}(E_e, \omega) = 
  \int\frac{d\Omega_{e\gamma}}{4\pi}
  \frac{d\Omega_{\nu}}{4\pi}\,\frac{1}{2}\sum_{\rm \lambda',
    pol}\frac{\omega^2 |{\cal M}(n \to p e^-
    \bar{\nu}_e\gamma)_{\lambda'}|^2}{(1 + 3 \lambda^2) 32 m^2_n E_e
    E_{\nu}}\,\Phi_{\beta\gamma}(\vec{k}_e,\vec{k},
  \vec{k}_{\nu}),
\end{eqnarray}
where $d\Omega_{e\gamma} = \sin\theta_{e\gamma}
d\theta_{e\gamma}d\varphi_{e\gamma}$ is an infinitesimal solid angle
of the electron--photon momentum correlations $\vec{k}_e\cdot \vec{n}
= k_e\cos \theta_{e\gamma}$, defined by the polar angle $0 \le
\theta_{e\gamma} \le \pi$ and the azimuthal angle $0 \le
\varphi_{e\gamma} \le 2\pi$, and $d\Omega_{\nu}$ is an infinitesimal
solid angle of the antineutrino momentum $\vec{k}_{\nu}$.  The
function $\Phi_{\beta\gamma}(\vec{k}_e,\vec{k}\,\vec{k}_{\nu})$ is
defined by the integral \cite{Ivanov2013,Ivanov2013a}.
\begin{eqnarray}\label{eq:A.35}
\Phi_{\beta\gamma}(\vec{k}_e,\vec{k}, \vec{k}_{\nu}) = \int^{\infty}_0
\delta(m_n - \sqrt{m^2_p + (\vec{k}_e + \vec{k} + \vec{k}_{\nu})^2} -
E_e - E_{\nu} - \omega)\,\frac{m_n}{E_p}\,\frac{E^2_{\nu}
  dE_{\nu}}{(E_0 - E_e - \omega)^2}.
\end{eqnarray}
The result of the integration over $E_{\nu}$ is equal to
\begin{eqnarray}\label{eq:A.36}
\Phi_{\beta\gamma}(\vec{k}_e,\vec{k}, \vec{k}_{\nu}) =
\frac{\displaystyle \Big(1 + \frac{1}{E_0 - E_e -
    \omega}\frac{k_e\cdot k}{~m_n}\Big)^2}{\displaystyle \Big(1 -
  \frac{1}{~m_n}\frac{(k_e + k)\cdot k_{\nu}}{E_{\nu}}\Big)^3},
\end{eqnarray}
where the antineutrino energy is defined by
\begin{eqnarray}\label{eq:A.37}
E_{\nu} = \frac{\displaystyle E_0 - E_e - \omega + \frac{k_e\cdot
    k}{~m_n}}{\displaystyle 1 - \frac{1}{~m_n}\,(E_e + \omega -
  |\vec{k}_e + \vec{k}\,|\cos\theta_{\nu})}.
\end{eqnarray}
Here $\theta_{\nu}$ is an angle between the momenta $\vec{k}_e +
\vec{k}$ and $\vec{k}_{\nu}$. To order $1/M$ the function
$\Phi_{\beta\gamma}(\vec{k}_e,\vec{k}, \vec{k}_{\nu})$ is given by
\begin{eqnarray}\label{eq:A.38}
\Phi_{\beta\gamma}(\vec{k}_e,\vec{k}, \vec{k}_{\nu}) = 1 +
\frac{2}{M}\,\frac{\omega}{E_{\nu}}\Big(E_e - \vec{k}_e\cdot
\vec{n}\,\Big) + \frac{3}{M}\Big(E_e + \omega - \frac{(\vec{k}_e +
  \omega \vec{n}\,)\cdot \vec{k}_{\nu}}{E_{\nu}}\Big),
\end{eqnarray}
where $E_{\nu} = E_0 - E_e - \omega$ \cite{Ivanov2013}. Setting
$\omega = 0$ we arrive at the corresponding function of the neutron
$\beta^-$--decay \cite{Ivanov2013}. For the calculation of the function
$\rho_{\beta\gamma}(E_e, \omega)$ we propose, first, to integrate over
the directions of the antineutrino 3--momentum. Now making the
integration in Eq.(\ref{eq:A.34}) over the antineutrino momentum solid
angle we arrive at the expression
\begin{eqnarray}\label{eq:A.39}
\hspace{-0.3in}&&\int \frac{d\Omega_{\nu}}{4\pi}\,\frac{1}{2}\sum_{\rm
  \lambda', pol}\frac{\omega^2 |{\cal M}(n \to p e^-
  \bar{\nu}_e\gamma)_{\lambda'}|^2}{(1 + 3 \lambda^2) 32 m^2_n E_e
  E_{\nu}}\,\Phi_{\beta\gamma}(\vec{k}_e,\vec{k}, \vec{k}_{\nu}) =
\Big[1 + 2\,\frac{\omega}{M}\,\frac{E_e - \vec{k}_e\cdot \vec{n}}{E_0
    - E_e - \omega} + \frac{3}{M}\,\Big(E_e + \omega -
  \frac{1}{3}\,E_0\Big) \nonumber\\
\hspace{-0.3in}&&+ \frac{\lambda^2 - 2(\kappa + 1)\lambda + 1}{1
  + 3\lambda^2}\,\frac{E_0 - E_e - \omega}{M}\Big]\,\Big[\Big(1 +
  \frac{\omega}{E_e}\Big)\,\frac{k^2_e - (\vec{k}_e\cdot
    \vec{n}\,)^2}{(E_e - \vec{k}_e\cdot \vec{n}\,)^2} +
  \frac{\omega^2}{E_e}\,\frac{1}{E_e - \vec{k}_e\cdot \vec{n}}\Big] +
\frac{3\lambda^2 - 1}{1 + 3 \lambda^2}\,\frac{1}{M}\nonumber\\
\hspace{-0.3in}&&\times\,\Big\{\frac{k^2_e + \omega \vec{k}_e\cdot
  \vec{n}}{E_e}\,\Big(\frac{k^2_e - (\vec{k}_e\cdot \vec{n}\,)^2}{(E_e
  - \vec{k}_e\cdot \vec{n}\,)^2} + \frac{\omega}{E_e - \vec{k}_e\cdot
  \vec{n}}\Big) + (\omega + \vec{k}_e\cdot \vec{n}\,)\Big[\Big(1 +
  \frac{\omega}{E_e}\Big)\frac{\omega}{E_e - \vec{k}_e\cdot \vec{n}} -
  \frac{m^2_e}{E_e}\,\frac{\omega}{(E_e - \vec{k}_e\cdot
    \vec{n}\,)^2}\Big]\Big\}\nonumber\\
\hspace{-0.3in}&& - \frac{\lambda^2 + 2 (\kappa + 1)\lambda
  - 1}{1 + 3\lambda^2}\,\frac{1}{M}\,\Big[\frac{k^2_e + \omega^2 +
    2\omega \vec{k}_e\cdot \vec{n}}{E_e}\,\frac{k^2_e -
    (\vec{k}_e\cdot \vec{n}\,)^2}{(E_e - \vec{k}_e\cdot \vec{n}\,)^2}
  + \frac{\omega}{E_e}\,\frac{k^2_e - (\vec{k}_e\cdot
    \vec{n}\,)^2}{E_e - \vec{k}_e\cdot \vec{n}} +
  \frac{\omega^2}{E_e}\,\frac{\omega + \vec{k}_e\cdot \vec{n}}{E_e -
    \vec{k}_e\cdot \vec{n}}\Big]\nonumber\\
\hspace{-0.3in}&& - \frac{\lambda(\lambda - 1)}{1 +
  3\lambda^2}\,\frac{1}{M}\,\Big[\frac{\omega}{E_e}\, \frac{k^2_e -
    (\vec{k}_e\cdot \vec{n}\,)^2}{E_e - \vec{k}_e\cdot \vec{n}} +
  3\,\frac{\omega^2}{E_e}\Big].
\end{eqnarray}
Plugging Eq.(\ref{eq:A.39}) into Eq.(\ref{eq:A.34}) we define the
function $\rho_{\beta\gamma}(E_e, \omega)$ by the integral over
electron--photon correlation angles only. It is given by
\begin{eqnarray}\label{eq:A.40}
\hspace{-0.3in}&&\rho_{\beta\gamma}(E_e, \omega) = \int
\frac{d\Omega_{e\gamma}}{4\pi}\, \Big[1 +
  2\,\frac{\omega}{M}\,\frac{E_e - \vec{k}_e\cdot \vec{n}}{E_0 - E_e -
    \omega} + \frac{3}{M}\,\Big(E_e + \omega - \frac{1}{3}\,E_0\Big) +
  \frac{\lambda^2 - 2(\kappa + 1)\lambda + 1}{1 +
    3\lambda^2}\,\frac{E_0 - E_e - \omega}{M}\Big]\nonumber\\
\hspace{-0.3in}&&\times\,\Big[\Big(1 +
  \frac{\omega}{E_e}\Big)\,\frac{k^2_e - (\vec{k}_e\cdot
    \vec{n}\,)^2}{(E_e - \vec{k}_e\cdot \vec{n}\,)^2} +
  \frac{\omega^2}{E_e}\,\frac{1}{E_e - \vec{k}_e\cdot \vec{n}}\Big] +
\frac{3\lambda^2 - 1}{1 + 3 \lambda^2}\,\frac{1}{M}\,\Big\{\frac{k^2_e
  + \omega \vec{k}_e\cdot \vec{n}}{E_e}\,\Big(\frac{k^2_e -
  (\vec{k}_e\cdot \vec{n}\,)^2}{(E_e - \vec{k}_e\cdot \vec{n}\,)^2} +
\frac{\omega}{E_e - \vec{k}_e\cdot \vec{n}}\Big)\nonumber\\
\hspace{-0.3in}&& + (\omega + \vec{k}_e\cdot \vec{n}\,)\Big[\Big(1 +
  \frac{\omega}{E_e}\Big)\frac{\omega}{E_e - \vec{k}_e\cdot \vec{n}} -
  \frac{m^2_e}{E_e}\,\frac{\omega}{(E_e - \vec{k}_e\cdot
    \vec{n}\,)^2}\Big]\Big\} - \frac{\lambda^2 + 2 (\kappa + 1)\lambda
  - 1}{1 + 3\lambda^2}\,\frac{1}{M}\,\Big[\frac{k^2_e + \omega^2 +
    2\omega \vec{k}_e\cdot \vec{n}}{E_e}\nonumber\\
\hspace{-0.3in}&& \times\,\frac{k^2_e - (\vec{k}_e\cdot
  \vec{n}\,)^2}{(E_e - \vec{k}_e\cdot \vec{n}\,)^2} +
\frac{\omega}{E_e}\,\frac{k^2_e - (\vec{k}_e\cdot \vec{n}\,)^2}{E_e -
  \vec{k}_e\cdot \vec{n}} + \frac{\omega^2}{E_e}\,\frac{\omega +
  \vec{k}_e\cdot \vec{n}}{E_e - \vec{k}_e\cdot \vec{n}}\Big] -
\frac{\lambda(\lambda - 1)}{1 +
  3\lambda^2}\,\frac{1}{M}\,\Big[\frac{\omega}{E_e}\, \frac{k^2_e -
    (\vec{k}_e\cdot \vec{n}\,)^2}{E_e - \vec{k}_e\cdot \vec{n}} +
  3\,\frac{\omega^2}{E_e}\Big].
\end{eqnarray}
To leading order in large proton mass expansion, i.e. at $M \to
\infty$, Eq.(\ref{eq:A.40}) reduces to the expression, calculated in
\cite{Ivanov2013}.

\end{document}